# Spectroscopic Imaging STM Studies of Electronic Structure in the Superconducting and Pseudogap Phases of Cuprate High-$T_c$ Superconductors

**Kazuhiro Fujita** [1,2,3], **Andrew R. Schmidt** [1,2,4], **Eun-Ah Kim** [1], **Michael J. Lawler** [1,5], **Dung Hai Lee** [4], **J. C. Davis** [1,2,6,7], **Hiroshi Eisaki** [8], **Shin-ichi Uchida** [3]

1. *LASSP, Department of Physics, Cornell University, Ithaca, NY 14853, USA*
2. *CMPMS Department, Brookhaven National Laboratory, Upton, NY 11973, USA*
3. *Department of Physics, University of Tokyo, Bunkyo-ku, Tokyo 113-0033, Japan*
4. *Department of Physics, University of California, Berkeley, California 94720, USA.*
5. *Department of Physics & Astronomy, Binghamton University, Binghamton, NY 13902, USA*
6. *School of Physics & Astronomy, U. of St. Andrews, St. Andrews, Fife KY16 9SS, Scotland.*
7. *Kavli Institute at Cornell for Nanoscale Science, Cornell University, Ithaca, NY 14850, USA.*
8. *Institute of Advanced Industrial Science and Technology, Tsukuba, Ibaraki 305-8568, Japan.*

One of the key motivations for the development of atomically resolved spectroscopic imaging STM (SI-STM) has been to probe the electronic structure of cuprate high temperature superconductors. In both the *d*-wave superconducting (dSC) and the pseudogap (PG) phases of underdoped cuprates, two distinct classes of electronic states are observed using SI-STM. The first class consists of the dispersive Bogoliubov quasiparticles of a homogeneous d-wave superconductor. These are detected below a lower energy scale $|E|=\Delta_0$ and only upon a momentum space (k-space) arc which terminates near the lines connecting k=$\pm(\pi/a_0,0)$ to k=$\pm(0,\pi/a_0)$. Below optimal doping, this 'nodal' arc shrinks continuously with decreasing hole density. In both the dSC and PG phases, the only broken symmetries detected in the $|E|\leq\Delta_0$ states are those of a d-wave superconductor. The second class of states occurs at energies near the pseudogap energy scale $|E|\sim\Delta_1$ which is associated conventionally with the 'antinodal' states near k=$\pm(\pi/a_0,0)$ and k=$\pm(0,\pi/a_0)$. We find that these states break the expected 90°-rotational (C$_4$) symmetry of electronic structure within CuO$_2$ unit cells, at least down to 180°-rotational (C$_2$) symmetry (nematic) but in a spatially disordered fashion. This intra-unit-cell C$_4$ symmetry breaking coexists at $|E|\sim\Delta_1$ with incommensurate conductance modulations locally breaking both rotational and translational symmetries (smectic). The characteristic wavevector Q of the latter is determined, empirically, by the k-space points where Bogoliubov quasiparticle interference terminates, and therefore evolves continuously with doping. The properties of these two classes of $|E|\sim\Delta_1$ states are indistinguishable in the dSC and PG phases. To explain this segregation of *k*-space into the two regimes distinguished by the symmetries of their electronic states and their energy scales $|E| \sim \Delta_1$ and $|E| \leq \Delta_0$, and to understand how this impacts the electronic phase diagram and the mechanism of high-$T_c$ superconductivity, represents one of a key challenges for cuprate studies.

## 1. Hole-doped Cuprates

The CuO$_2$ plane electronic structure is dominated by Cu *3d* and O *2p* orbitals.[1] Each Cu $d_{x2-y2}$ orbital is split energetically into singly and doubly occupied configurations by on-site Coulomb interactions. This results in a 'charge-transfer' Mott insulator[1] that is also strongly antiferromagnetic due to inter-copper superexchange.[2,3] 'Hole-doping' is achieved by removing electrons from the O atoms.[4,5] The phase diagram[6] as a function of $p$, the number of holes per CuO$_2$, is shown schematically in Fig. 1a. Antiferromagnetism exists for $p < 2$-5%, superconductivity occurs in the range 5-10% $< p <$ 25-30%, and a Fermi liquid state appears for $p > 25$-30%. The highest superconducting critical temperature $T_c$ occurs at so-called 'optimal' doping $p \sim 16$% and the superconductivity exhibits d-wave symmetry.

With reduced $p$, an unusual electronic excitation with energy scale $|E| = \Delta_1$, and which is anisotropic in **k**-space,[6,7,8,9,10,11] appears at $T^* > T_c$. This region of the phase diagram has been labelled the 'pseudogap' (PG) phase because the energy scale $\Delta_1$ could be the energy gap of a distinct electronic phase.[7,8] Explanations for the PG phase include (i) that it occurs due to hole-doping an antiferromagnetic Mott insulator to create a spin-liquid[3,12,13,14,15,16] or, (ii) that it is a d-wave superconductor lacking long range phase coherence[17,18,19,20,21,22] or, (iii) that it is a distinct electronic ordered phase.[23,24,25,26,27,28,29,30,31,32,33,34,35,36,37] It has not yet been possible to determine whether one of these proposals (or some combination thereof) is correct, or what the microscopic influence of the PG phase is upon the dSC phase.

Two energy scales $\Delta_1$ and $\Delta_0$ associated with two distinct types of electronic excited states[6,7,8,9,38,39,40,41] are observed in underdoped cuprates by multiple distinct spectroscopies, and $\Delta_0$ and $\Delta_1$ diverge from one another with diminishing $p$ (Fig. 1b reproduced from ref.8). For example, optical transient grating spectroscopy reveals that the $|E| \sim \Delta_1$ excitations propagate very slowly without recombination into Cooper pairs, whereas the lower energy 'nodal' excitations propagate and reform delocalized Cooper pairs as expected.[38] Andreev tunneling also finds two distinct excitation energy scales which diverge as $p \rightarrow 0$: the first is identified with the pseudogap energy $\Delta_1$ and the second lower scale $\Delta_0$ with the maximum pairing gap energy of delocalized Cooper pairs.[39] Raman spectroscopy indicates that only the scattering near the d-wave node is consistent with delocalized Cooper pairing.[40] The superfluid density measured using muon spin rotation evolves with hole-density in a manner inconsistent the whole Fermi surface being available for delocalized Cooper pairing.[41]

Figure 1c shows a schematic depiction of the Fermi surface within the $CuO_2$ Brillouin zone, and distinguishes the 'nodal' from 'antinodal' regions of **k**-space. Momentum-resolved examination of cuprate electronic structure using angle resolved photoemission spectroscopy (ARPES) in the PG phase reveals that excitations with $E \sim -\Delta_1$ occur near the antinodal regions $\mathbf{k} \cong (\pi/a_0, 0); (0, \pi/a_0)$, and that $\Delta_1(p)$ increases rapidly as $p \rightarrow 0$.[7,8,9,10] By contrast, the nodal region of **k**-space exhibits an ungapped 'Fermi Arc'[42] in the PG phase, and a momentum- and temperature-dependent energy gap opens upon this arc in the dSC phase.[42,43,44,45,46,47,48]

Density-of-states measurements from tunneling spectroscopy report an energetically particle-hole symmetric excitation energy $|E| = \Delta_1$ that is unchanged in the PG and dSC phases.[49, 50] Figure 2(b) shows the evolution of spatially-averaged differential tunnelling conductance $g(E)$ for $Bi_2Sr_2CaCu_2O_{8+\delta}$.[51,52,53] The $p$-dependence of the pseudogap energy $E = \pm \Delta_1$ is indicated with blue dashed curve while that of $\Delta_0$ (as discussed in § 6 is shown by red dashed curves. SI-STM has amplified upon these observations by using atomically resolved and registered tunneling to visualize the distinct spatial structure of both types of states. For energies $|E| \leq \Delta_0$, the dispersive Bogoliubov quasiparticles (Fig. 2(a)) of a spatially homogeneous superconductor are always observed.[54,55,56,57,58,59,60] The states near $|E| \sim \Delta_1$ are, in contrast, spatially disordered on the nanometer scale[51,52,53,61,62,63,64,65,66,67,68] and their spatial structure exhibits several distinct broken symmetries (Fig. 2(c)).[51,58,59,60,69,70]

## 2.    $Bi_2Sr_2CaCu_2O_8$ Samples and Preparation

High-quality single crystals of $Bi_2Sr_2CaCu_2O_8$ were synthesized for our studies using the traveling-solvent-floating-zone (TSFZ) method. The samples are of $Bi_{2.1}Sr_{1.9}CaCu_2O_{8+\delta}$ and $Bi_{2.2}Sr_{1.8}Ca_{0.8}Dy_{0.2}Cu_2O_{8+\delta}$ and were synthesized from well-dried powders of $Bi_2O_3$,

SrCO₃, CaCO₃, Dy₂O₃, and CuO. The crystal growth was carried out in air and at growth speeds of $0.15 - 0.2$ mm/h for all the samples.

Inductively coupled plasma spectroscopy was used for the composition analysis and a superconducting quantum interference device (SQUID) magnetometer was used for measurement of $T_c$. $T_c$ is defined as the onset temperature at which the zero-field-cooled susceptibility starts to drop. Annealing is used to vary the critical temperature of each sample. Oxidation annealing is performed in air or under oxygen gas flow, and deoxidation annealing is done in vacuum or under nitrogen gas flow for the systematic study at different hole-densities.[71]

We have studied sequence of $Bi_2Sr_2CaCu_2O_{8+\delta}$ samples for which $p \cong 0.19, 0.17, 0.14,$ 0.12, 0.10, 0.08, 0.06 or with $T_c$(K) = 86, 88, 74, 64, 45, 37, 20 respectively, with many these samples being studied in both the dSC and PG phases.[51,52,53,54,55,56,58,59,60,61,62,69,70] Each sample is inserted into the cryogenic ultra high vacuum of the SI-STM system,[72] cleaved to reveal an atomically clean BiO surface, and all measurements were made between 1.9 K and 65 K. Three cryogenic SI-STM's were used during these studies and the resulting data consists of >10⁸ atomically resolved and registered tunneling spectra measured at the BiO surface of $Bi_2Sr_2CaCu_2O_{8+\delta}$.

## 3.    Spectroscopic Imaging Scanning Tunneling Microscopy

Spectroscopic imaging STM consists of measuring the tip-sample differential tunneling conductance $dI/dV(\mathbf{r}, E = eV) \equiv g(\mathbf{r}, E = eV)$ with atomic resolution and register and as a function of both location $\mathbf{r}$ and electron energy $E$. It is distinct from other electron spectroscopies in that it can access simultaneously the real space ($\mathbf{r}$-space) and momentum space ($\mathbf{k}$-space) electronic structure for states both above and below the Fermi level. There are, however, common systematic errors that become especially prevalent and significant in studies of underdoped cuprates.

The first and most elementary issue emerges from the tunneling current equation

$$I(\mathbf{r}, z, V) = f(\mathbf{r}, z) \int_0^{eV} N(\mathbf{r}, E)\, dE \qquad (1)$$

where $z$ is the tip-surface distance, $V$ the tip-sample bias voltage, $N(\mathbf{r}, E)$ the sample's local-density-of-electronic-states, while $f(\mathbf{r}, z)$ contains effects of tip elevation and of tunneling matrix elements. The $g(\mathbf{r}, E)$ data are then related to $N(\mathbf{r}, E)$ by [56, 58,59,60, 61]

$$g(\mathbf{r}, E = eV) = \frac{eI_S}{\int_0^{eV_S} N(\mathbf{r}, E')\,dE'} N(\mathbf{r}, E) \tag{2}$$

where $V_S$ and $I_S$ are the (constant but arbitrary) junction 'set-up' bias voltage and current respectively. From eq. (2) we see that when $\int_0^{eV_S} N(\mathbf{r}, E')\,dE'$ is strongly heterogeneous at the atomic scale as in underdoped cuprates,[51,52,53,57,58,59,60,61,62,63,64,65,66,67,68,69] $g(\mathbf{r}, E = eV)$ cannot be used to measure the spatial arrangements of $N(\mathbf{r}, E)$ because the denominator if unknown and wildly fluctuating.[58] Mitigation[56,58,59,60] of these potentially severe systematic errors can be achieved by using either

$$Z(\mathbf{r}, E) \equiv \frac{g(\mathbf{r}, E = +eV)}{g(\mathbf{r}, E = -eV)} = \frac{N(\mathbf{r}, +E)}{N(\mathbf{r}, -E)} \tag{3}$$

or the related but non-energy-resolved

$$R(\mathbf{r}) \equiv \frac{I(\mathbf{r}, E = +eV)}{I(\mathbf{r}, E = -eV)} = \frac{\int_0^{+eV} N(\mathbf{r}, E)\,dE}{\int_{-eV}^0 N(\mathbf{r}, E)\,dE} \tag{4}$$

The observables in eqs. (3) and (4) then allow distances, wavelengths and symmetries to be measured correctly but at the expense of mixing information derived from states at $\pm E$.

A different more specific challenge is the random nanoscale variation of $\Delta_1(\mathbf{r})$ which causes the $|E| \sim \Delta_1$ pseudogap states to be detected at different locations for different bias voltages (Fig. 5(a)). This problem can be lessened[59, 60, 69,70] by scaling the tunnel-bias energy $E=eV$ at each $\mathbf{r}$ by the pseudogap magnitude $\Delta_1(\mathbf{r})$ at the same location. This procedure defines a reduced energy scale $e = E/\Delta_1(\mathbf{r})$ such that

$$Z(\mathbf{r}, e) \equiv Z(\mathbf{r}, E/\Delta_1(\mathbf{r})) \tag{5}$$

in which the $|E| \sim \Delta_1$ states all occur together at $e=1$.[59]

Another important systematic error occurs when using $g(\mathbf{q}, E)$ and $Z(\mathbf{q}, E)$, the power spectral density Fourier transforms of $g(\mathbf{r}, E)$ and $Z(\mathbf{r}, E)$ respectively. To achieve sufficient precision in $|\boldsymbol{q}(E)|$ for discrimination of a non-dispersive ordering wavevector $\boldsymbol{Q}^*$ due to an electronic ordered phase, from the dispersive wavevectors $\boldsymbol{q}(E)$ due to quantum interference patterns of delocalized states, requires that $g(\mathbf{r},E)$ or $Z(\mathbf{r},E)$ be measured in large fields-of-view (FOV) and with energy resolution at or below ~2 meV. Using a smaller FOV or poorer energy resolution in $g(\mathbf{r},E)$ studies generates inexorably the erroneous impression of non-dispersive modulations. For $Bi_2Sr_2CaCu_2O_{8+\delta}$ in both the dSC and PG phases, no deductions distinguishing between dispersive and non-dispersive excitations can be made using Fourier transformed $g(\mathbf{r},E)$ data from a FOV smaller than ~45nm-square.[55, 60]

A final systematic error derives from the slow picometer scale drifts of the tip location due to both mechanical creep of the piezoelectric actuators and mK temperature variations, over the continuous and extended period of up to a week required for each $g(\mathbf{r},E)$ data set to be acquired. This is particularly critical in research requiring a precise knowledge of the spatial phase of the Cu lattice.[69,70] We have introduced a post-measurement partial correction for these effects that uses the identification of a slowly varying "displacement" field $\vec{u}(\mathbf{r})$ such that the corrected positions $\mathbf{r} - \vec{u}(\mathbf{r})$ will form a perfect square lattice for the sites of Bi (Cu) atoms ($\vec{d}_{Cu} = 0$ within each $CuO_2$ unit cell). To achieve this, we considered topographic images taking the form

$$T(\mathbf{r}) = T_0\Big(\cos\big(\mathbf{Q}_x \cdot (\mathbf{r} - \vec{u}(\mathbf{r}))\big) + \cos\big(\mathbf{Q}_y \cdot (\mathbf{r} - \vec{u}(\mathbf{r}))\big)\Big) + T_{sup}\cos\big(\mathbf{Q}_{sup} \cdot (\mathbf{r} - \vec{u}(\mathbf{r}))\big) + \dots \quad (6)$$

where $\mathbf{Q}_{sup}$ refers to the crystalline supermodulation. That $\vec{u}(\mathbf{r})$ is slowly varying compared to the scale of the lattice is verified from the relative sharpness of Bragg peaks $\pm\vec{Q}_{x,y}$ in $T(\mathbf{q})$, the PSD Fourier transform of the topograph. To extract $\vec{u}(\mathbf{r})$, coarsening length scale $1/\Lambda_u$ is introduced over which $\vec{u}(\mathbf{r})$ is roughly constant such that $\Lambda_u << \big|\mathbf{Q}_{sup}\big|, \big|\vec{Q}_{x,y}\big|$. Analysis of $T(\mathbf{q})$ itself determines when one can safely choose a fairly small $\Lambda_u$ because, in that case, Bragg peaks are quite sharp. Next we consider

$$T_x(\mathbf{r}) = \sum_{\mathbf{r}'} T(\mathbf{r}')e^{-i\mathbf{Q}_x \cdot \mathbf{r}'}\left(\frac{\Lambda_u^2}{2\pi}e^{-\Lambda_u^2|\mathbf{r}-\mathbf{r}'|^2/2}\right) \quad (7)$$

the weighted average of $T(\mathbf{r}')e^{-i\mathbf{Q}_x \cdot \mathbf{r}'}$ over the length scale $1/\Lambda_u$. Since $\Lambda_u << \big|\mathbf{Q}_{sup}\big|, \big|\mathbf{Q}_{x,y}\big|$ their contributions average out, leaving

$$T_x(\mathbf{r}) \approx (T_0/2)e^{-i\mathbf{Q}_x \cdot \vec{u}(\mathbf{r})} \quad (8)$$

because $\vec{u}(\mathbf{r}') \approx \vec{u}(\mathbf{r})$ for small $|\mathbf{r}-\mathbf{r}'| < \Lambda_u$. Similarly

$$T_y(\mathbf{r}) = \sum_{\mathbf{r'}} T(\mathbf{r'}) e^{-i\mathbf{Q}_y \cdot \mathbf{r'}} \left( \frac{\Lambda_u^2}{2\pi} e^{-\Lambda_u^2 |\mathbf{r} - \mathbf{r'}|^2 / 2} \right) \approx (T_0/2) e^{-i\mathbf{Q}_y \cdot \vec{u}(\mathbf{r})} \qquad (9)$$

Hence one can estimate $\vec{u}(\mathbf{r})$ and thus, by inverting all the distortion-induced displacements in the raw $T(\mathbf{r})$ data, undo effects of piezoelectric and/or thermal drift and cause the topographic image to become perfectly periodic. This same geometrical transformations to undo $\vec{u}(\mathbf{r})$ is then carried out on each $g(\mathbf{r},E)$ acquired simultaneously with the $T(\mathbf{r})$, so that the processed $T(\mathbf{r})$ and $g(\mathbf{r},E)$ are then registered to each other and to a perfectly $a_0$ periodic square lattice.

## 4.  Effect of Non-Magnetic and Magnetic Impurity Atoms

Substitution of magnetic and non-magnetic impurity atoms can be used to probe the microscopic electronic structure of an unconventional superconductor, and especially whether there are sign changes on the order parameter.[73,74,75,76] For a superconductor describable by BCS theory, if the order parameter exhibits S-wave symmetry, then non-magnetic impurity atoms should have little effect because time reversed pairs of states which can undergo Cooper pairing are not disrupted. Magnetic impurity atoms, on the other hand should be quite destructive since they break time reversal symmetry. For unconventional superconductors (non S-wave) this simple situation does not pertain and both magnetic and potential scattering impurities produce strong pair breaking effects. However, the spatial/energetic structure of the bound and resonant states[73, 77] (which are produced by Bogoliubov quasiparticle scattering at impurity atoms) can be highly revealing of the microscopic order parameter symmetry. These theoretical ideas as summarized in ref. 73 were the basis for SI-STM studies of non-magnetic Zn impurity atoms and magnetic Ni impurity atoms substituted on the Cu sites of $Bi_2Sr_2CaCu_2O_{8+\delta}$ .[78,79]

For Zn-doped $Bi_2Sr_2CaCu_2O_{8+\delta}$ near optimal doping, a typical $g(\mathbf{r},E=eV)$ of a 50nm square region at $V$=-1.5 mV is shown in Fig. 3(a) with the overall dark background being indicative of a very low $g(\mathbf{r},E)$ near the Fermi level. This is as expected for a superconductor far below $T_c$. However there are a number of randomly distributed bright sites corresponding to areas of high $g(\mathbf{r},E)$, each with a distinct four-fold symmetric shape and the same relative orientation. In Fig. 3(b) we show a comparison between spectra taken exactly at their centers and spectra taken at usual superconducting regions of the sample. The spectrum at the center of a bright site has a very strong intra-gap conductance peak at energy $\Omega$=-1.5±0.5 meV. And, at these sites, the superconducting coherence peaks (identified by the arrows in Fig. 3(c)) are strongly diminished, indicating the suppression of superconductivity. All of these phenomena are among the theoretically predicted characteristics of a very strong (almost unitary) quasiparticle scattering resonance at a single potential-scattering impurity atom in a *d*-wave superconductor.[73]

Studies of Ni-doped near optimal doping revealed more intriguing results. As an example, Fig. 4 shows two simultaneously acquired $g(\mathbf{r},E=eV)$ maps taken on Ni-doped $Bi_2Sr_2CaCu_2O_{8+\delta}$ at sample bias $V_{bias}$=±10mV. They reveal both the particle-like (positive bias) and hole-like (negative bias) components of one of the impurity states that exist at each Ni. At +10mV `+-shaped' regions of higher $g(\mathbf{r},E)$ are observed, whereas at -10mV the corresponding higher $g(\mathbf{r},E)$ regions are `X-shaped'. $g(\mathbf{r},E)$ maps at $V_{bias}$=±19mV show the particle-like and hole-like components of a second impurity state at Ni whose spatial structure is very similar to that at $V_{bias}$=±10 mV. Figure 4(c) shows the typical spectra taken at the Ni atom site in which there are two clear particle-like $g(\mathbf{r},E)$ peaks. The average magnitudes of these on-site impurity-state energies are $\Omega_1$=9.2±1.1meV and $\Omega_2$=18.6±0.7 meV. The existence of two states is as expected for a magnetic impurity in a d-wave superconductor.[73] Perhaps most significant, however, is the fact that the magnetic impurity does not appear to suppress the superconductivity (as judged by the coherence peaks) at all, as if magnetism is not anathema to the pairing interaction locally at atomic scale. This is not as expected within BCS-based models of the pairing mechanism.

Calculation of the potential scattering phase shift $\delta_0$=tan$^{-1}(\pi N_F U)$ for Ni gives $\delta_0$= 0.36$\pi$, whereas Zn is a unitary scatterer ($\delta_0 \sim \pi/2$)[78] ($N_F$ the normal density of states per site at the Fermi energy, $U$ the strength of the potential scattering represented by on-site coulomb energy. $N_F U$=-0.67[79]). The similarity of these phase shifts imply that phenomena dependent on scattering should be quite similar in Ni- and Zn-doped samples. In fact, using these parameters in an Abrikosov-Gorkov model (and ignoring Ni's magnetic potential), we calculate that $T_c$ would be suppressed only about 20% faster by Zn than by Ni, certainly within the range of experimental observations.[80,81] This means that the understanding of potential scattering aspects of the impurity atoms in this d-wave superconductor is quite satisfactory.

One of the most interesting observations made during these impurity atom studies, and one which was not appreciated at the time of the original experiments, was that the vivid, clear and theoretically reasonable impurity states at Zn and Ni disappear as hole density $p$ is reduced below optimal doping in $Bi_2Sr_2CaCu_2O_{8+\delta}$.[62, 82, 83] Thus, even though the density of Zn or Ni impurity atoms is the same, the response of the electronic structure to them is quite different. In fact, the Zn and Ni impurity states (Fig. 3 and 4) quickly diminish in intensity and eventually become undetectable at low hole-density.[82,83] One possible explanation for this strong indication of anomalous electronic structure in underdoped $Bi_2Sr_2CaCu_2O_{8+\delta}$ could be that the $\textbf{\textit{k}}$-space states which contribute to Cooper pairing on the whole Fermi surface at optimal doping, no longer do so at lower $p$ (§ 6). In this situation, all the Bogoliubov eigenstates necessary for scattering resonances to be created[73,77] would no longer be available. This hypothesis is quite consistent with the discovery of restricted regions of $k$-space supporting coherent Bogoliubov quasiparticles that diminish in area with falling hole-density[59] as discussed in § 6.

## 5.    Nanoscale Electronic Disorder in $Bi_2Sr_2CaCu_2O_{8+\delta}$

Nanoscale electronic disorder is pervasive in images of $\Delta_1(\textbf{r})$ measured on $Bi_2Sr_2CaCu_2O_{8+\delta}$ samples.[51,52,53,55,58,59,60,61,62,63,64,65,66,67,68,69] The values of $|\Delta_1|$ range from above 130 meV to below 10 meV as $p$ ranges from 0.06 to 0.22. Similar nanoscale $\Delta_1(\textbf{r})$ disorder is seen in $Bi_2Sr_2CuO_{6+\delta}$[57, 66] and in $Bi_2Sr_2Ca_2Cu_3O_{10+\delta}$.[84] Figure 5(a) shows a typical $Bi_2Sr_2CaCu_2O_{8+\delta}$ $\Delta_1(\textbf{r})$ image - upon which the sites of the non-stoichiometric oxygen dopant ions are overlaid as white dots.[52] Figure 5(b) shows the typical $g(E)$ spectrum associated with each different value of $\pm \Delta_1$.[51] It also reveals quite vividly how the electronic structure becomes homogeneous[51,52,53,56,57,59,60] for $|E| \leq \Delta_o$ as indicated by the arrows. Samples of $Bi_2Sr_2CuO_{6+\delta}$ and $Bi_2Sr_2Ca_2Cu_3O_{10+\delta}$ show virtually identical effects.[57,66,84] Moreover imaging $\Delta_1(\textbf{r})$ in the PG phase reveals highly similar[60, 65, 66, 67] nanoscale electronic disorder.

A key component of the explanation for these phenomena is that electron-acceptor atoms must be introduced[85] to generate hole doping. This almost always creates random distributions of differently charged dopant ions near the $CuO_2$ planes.[86] The dopants in $Bi_2Sr_2CaCu_2O_{8+\delta}$ are -2$e$ oxygen ions charged interstitials and may cause a range of different local effects. For example, electrostatic screening cause holes to congregate surrounding the dopant locations thereby reducing the energy-gap values nearby.[87, 88] Or the dopant ions could cause nanoscale crystalline stress/strain[89,90,91,92,93] thereby disordering hopping matrix elements and electron-electron interactions within the $CuO_2$ unit cell. In $Bi_2Sr_2CaCu_2O_{8+\delta}$ the locations of interstitial dopant ions are identifiable because an electronic impurity state occurs at E=-0.96V nearby each ion (Fig. 5(a)).[52] Strong spatial correlations are observed between the distribution of these impurity states and $\Delta_1(\mathbf{r})$ maps. This implies that dopant ion disorder is responsible for much of the $\Delta_1(\mathbf{r})$ electronic disorder. The principal effect near each dopant is a shift of spectral weight from low to high energy, with $\Delta_1$ increasing strongly. Simultaneous imaging of the dopant ion locations and $g(\mathbf{r},E<\Delta_0)$ reveals that the dispersive $g(\mathbf{r},E)$ modulations due to scattering of Bogoliubov quasiparticles are well correlated with dopant ion locations meaning that the dopant ions are an important source of such scattering (§ 6).[51,52,54,55,56,57,59,60] Thus, it is the chemical doping process itself which both disorders $\Delta_1$ and causes quasiparticle scattering is important because similar nanoscale electronic disorder phenomena must occur commonly in all non-stoichiometric cuprates.

The microscopic mechanism of the $\Delta_1$-disorder is not yet fully understood. Hole-accumulation surrounding negatively charged oxygen dopant ions does not appear to be the explanation because the modulations in integrated density of filled states are observed to be weak.[52] More significantly, $\Delta_1$ is actually strongly increased nearby the dopant ions[52] that is diametrically opposite to the expected effect from hole-accumulation there. Atomic substitution at random on the Sr site by Bi or by some other trivalent lanthanoid is known to suppress superconductivity strongly[86,94] possibly due to geometrical distortions of the unit cell and associated changes in the hopping matrix elements. It has therefore been proposed that the interstitial dopant ions might act similarly, perhaps by displacing the Sr or apical oxygen atoms[86,89,90,94] and thereby distorting the unit cell geometry. Direct support for this point of view comes from the observation that quasi-periodic distortions of the crystal unit-cell geometry yield virtually identical perturbations in $g(E)$ and $\Delta_1(\mathbf{r})$ but now are unrelated to the dopant ions.[95] Thus it seems that the $\Delta_1$-disorder is not caused primarily by carrier density modulations but by geometrical distortions to the unit cell dimensions with resulting strong local changes in the high energy electronic structure. One could also expect the presence of such disorder in $Ca_{2-x}Na_xCuO_2Cl_2$ as Ca is substituted by Na since and indeed similar $\Delta_1$ disorder is also observed in this material.[58]

Underdoped cuprate $g(E)$ spectra always exhibit "kinks"[51,52,53,56,57, 59,61,62,63,64,66,67,68] close to the energy scale where electronic homogeneity is lost. They are weak perturbations to $N(E)$ near optimal doping, becoming more clear as $p$ is diminished.[51,53] Figure 5(b) demonstrates how, in $\Delta_1$-sorted $g(E)$ spectra, the kinks are universal but become more obvious for $\Delta_1 > 50\,\text{meV}$.[51,53] Each kink can be identified and its energy is labelled by $\Delta_0(\mathbf{r})$. By determining $\overline{\Delta}_0$ (the spatial average of $\Delta_0(\mathbf{r})$) as a function of $p$, we find that this energy $\overline{\Delta}_0$ always divides the electronic structure into two categories.[53] For $|E| < \overline{\Delta}_0$ the excitations are homogenous in $\mathbf{r}$-space and well defined Bogoliubov quasiparticle eigenstates in $\mathbf{k}$-space (§ 6). By contrast, the pseudogap excitations at $|E| \sim \Delta_1$ are heterogeneous in $\mathbf{r}$-space and ill defined in $\mathbf{k}$-space (§ 7). Figure 5(c) provides a summary of the evolution of $\overline{\Delta}_0$ and $\overline{\Delta}_1$ with $p$.

To summarize: the $\Delta_1$-disorder of $Bi_2Sr_2CaCu_2O_{8+\delta}$ is strongly influenced by the random distribution of dopant ions.[52] It occurs through an electronic process in which geometrical distortions of the crystal unit cell play a prominent role.[91-95] The disorder is most strongly reflected in the states near the pseudogap energy $|E| \sim \Delta_1$. However, the states with $|E| \leq \Delta_0$ are homogeneous when studied using direct imaging[51,52,53,62] or using from quasiparticle interference as described in § 6.

## 6. Bogoliubov Quasiparticle Interference Imaging

Bogoliubov quasiparticle interference (QPI) occurs when quasiparticle de Broglie waves are scattered by impurities and the scattered waves undergo quantum interference. In a $d$-wave superconductor with a single hole-like band of uncorrelated electrons as sometimes used to describe $Bi_2Sr_2CaCu_2O_{8+\delta}$, the Bogoliubov quasiparticle dispersion $E(\mathbf{k})$ would exhibit constant energy contours which are 'banana-shaped'. The $d$-symmetry superconducting energy gap would then cause strong maxima to appear for a given $E$, in the joint-density-of-states at the eight banana-tips $\mathbf{k}_j(E)$; $j = 1, 2,..., 8$. Elastic scattering between the $\mathbf{k}_j(E)$ should produce $\mathbf{r}$-space interference patterns in $N(\mathbf{r},E)$. The resulting $g(\mathbf{r},E)$ modulations should exhibit 16 $\pm\mathbf{q}$ pairs of dispersive wavevectors in $g(\mathbf{q},E)$ (Fig. 6(a)). The set of these wavevectors characteristic of $d$-wave superconductivity consists of seven: $\mathbf{q}_i(E)$ $i=1,....,7$ with $\mathbf{q}_i(-E) = \mathbf{q}_i(+E)$. By using the point-group symmetry of the first $CuO_2$ Brillouin zone and this 'octet model',[96,97,98] the locus of the banana tips $\mathbf{k}_B(E) = (k_x(E),k_y(E))$ can be determined from:

$$\mathbf{q}_1 = (2k_x, 0) \qquad \mathbf{q}_4 = (2k_x, 2k_y) \qquad \mathbf{q}_7 = (k_x - k_y, k_y - k_x)$$
$$\mathbf{q}_2 = (k_x + k_y, k_y - k_x) \qquad \mathbf{q}_5 = (0, 2k_y) \tag{10}$$
$$\mathbf{q}_3 = (k_x + k_y, k_y + k_x) \qquad \mathbf{q}_6 = (k_x - k_y, k_y + k_x)$$

The $\mathbf{q}_i(E)$ are measured from $Z(\mathbf{q}, E)$, the Fourier transform of spatial modulations seen in $Z(\mathbf{r}, E)$ (see Fig. 2(a) for example), and the $\mathbf{k}_B(E)$ are then determined by using eq. (10) within the requirement that all its independent solutions be consistent at all energies. The superconductor's Cooper-pairing energy gap $\Delta(\mathbf{k})$ is then determined directly by inverting $\mathbf{k}_B(E = \Delta)$.

Near optimal doping in $Bi_2Sr_2CaCu_2O_{8+\delta}$, measurements from QPI of $\mathbf{k}_B(E)$ and $\Delta(\mathbf{k})$ (Fig. 6(b)) are consistent with ARPES.[55,99] And, in both $Ca_{2-x}Na_xCuO_2Cl_2$ and $Bi_2Sr_2Cu_1O_{6+\delta}$, the octet model yields $\mathbf{k}_B(E)$ and $\Delta(\mathbf{k})$ equally well.[56,57] Moreover, the basic validity of the fundamental $\mathbf{k}$-space phenomenology behind the d-wave QPI 'octet' model has been confirmed by ARPES studies.[100,101,102] Therefore, Fourier transformation of $Z(\mathbf{r}, E)$ in combination with the octet model of $d$-wave Bogoliubov QPI yields the two branches of the Bogoliubov excitation spectrum $\mathbf{k}_B(\pm E)$ plus the superconducting energy gap magnitude $\pm\Delta(\mathbf{k})$ along the specific $\mathbf{k}$-space trajectory $\mathbf{k}_B$ for both filled and empty states in a single experiment. And, since only the Bogoliubov states of a $d$-wave superconductor could exhibit such a set of 16 pairs of interference wavevectors with $\mathbf{q}_i(-E) = \mathbf{q}_i(+E)$ and all dispersions internally consistent within the octet model, the energy gap $\pm\Delta(\mathbf{k})$ determined by these procedures is definitely that of the delocalized Cooper-pairs.

We used these Bogoliubov QPI imaging techniques to study the evolution of $\mathbf{k}$-space electronic structure with diminishing $p$ in $Bi_2Sr_2CaCu_2O_{8+\delta}$. In the SC phase, the expected 16 pairs of $\mathbf{q}$-vectors are always observed in $Z(\mathbf{q}, E)$ and are found consistent with each other within the octet model (Fig. 2(a)). However, in underdoped $Bi_2Sr_2CaCu_2O_{8+\delta}$ the dispersion of octet model $\mathbf{q}$-vectors always stops at the same weakly doping-dependent[51,57,59] excitation energy $\Delta_0$ and at $\mathbf{q}$-vectors indicating that the relevant $\mathbf{k}$-space states are still far from the boundary of the Brillouin zone (Fig. 6(c)). These observations are quite unexpected in the context of the $d$-wave BCS octet model. Moreover, for $|E| > \Delta_0$ the dispersive octet of $\mathbf{q}$-vectors disappears and three ultra-slow dispersion $\mathbf{q}$-vectors become predominant. They are the reciprocal lattice vector $\mathbf{Q}$ along with $\mathbf{q}_1^*$ and $\mathbf{q}_5^*$ (see Fig. 6(a)). The ultra-slow dispersion incommensurate modulation wavevectors equivalent to $\mathbf{q}_1^*$ and $\mathbf{q}_5^*$ has also been detected by SI-STM in $Ca_{2-x}Na_xCuO_2Cl_2$[56] and $Bi_2Sr_2Cu_1O_{6+\delta}$,[57] and by ARPES in $Ca_{2-x}Na_xCuO_2Cl_2$[43] and $Bi_2Sr_2CaCu_2O_{8+\delta}$.[101,102]

We show in Fig. 6(c) the locus of Bogoliubov quasiparticle states $k_B(E)$ determined as a function of $p$ using QPI. Here discovered a quite surprising fact: when the Bogoliubov QPI patterns disappear at $\Delta_0$, the $k$-states are near the diagonal lines between $k=(0,\pi/a_0)$ and $k=(\pi/a_0,0)$ within the $CuO_2$ Brillouin zone. These $k$-space Bogoliubov arc tips are defined by both the change from clearly dispersive states to those whose dispersion is extremely slow or non existent, and by the disappearance of the $q_2$, $q_3$, $q_6$ and $q_7$ modulations. Thus, the QPI signature of delocalized Cooper pairing is confined to an arc (fine solid lines in Figure 6c) and this arc shrinks with falling $p$.[59] This observation has been supported directly by ARPES studies[41,48] and by QPI studies,[56, 57] and indirectly by analyses of $g(\mathbf{r},E)$ by fitting to a multi-parameter model for $k$-space structure of a dSC energy gap.[68]

The minima (maxima) of the Bogoliubov bands $k_B(\pm E)$ should occur at the $k$-space location of the Fermi surface of the non-superconducting state. One can therefore ask if the carrier-density count satisfies Luttinger's theorem, which states that twice the $k$-space area enclosed by the Fermi surface, measured in units of the area of the first Brillouin zone, equals the number of electrons per unit cell, $n$. In Fig. 6(c) we show as fine solid lines hole-like Fermi surfaces fitted to our measured $k_B(E)$. Using Luttinger's theorem with these $k$-space contours extended to the zone face would result in a calculated hole-density $p$ for comparison with the estimated $p$ in the samples. These data are shown by filled symbols in the inset to Figure 6c showing how the Luttinger theorem is violated at all doping below $p \sim 10\%$ if the large hole-like Fermi surface persists in the underdoped region of the phase diagram.

Figure 7 provides a doping-dependence analysis of the locations of the ends of the arc-tips at which Bogoliubov QPI signature disappears and where the $q_1^*$ and $q_5^*$ modulations appear. Figure 7(a) shows a typical $Z(\mathbf{r},E \sim \Delta_1)$ and its $Z(\mathbf{q},E)$ as an inset. Here the vectors $q_1^*$ and $q_5^*$ are labeled along with the Bragg vectors $\mathbf{Q}_x$ and $\mathbf{Q}_y$. Figure 7(b) shows the doping dependence for $Bi_2Sr_2CaCu_2O_{8+\delta}$ of the location of both $q_1^*$ and $q_5^*$ measured from $Z(\mathbf{q},E)$.[59] The measured magnitude of $q_1^*$ and $q_5^*$ versus $p$ are then shown in Fig. 7(c) along with the sum $q_1^*+q_5^*$ which is always equal to $2\pi$. This demonstrates that, as the Bogoliubov QPI extinction point travels along the line from $k=(0,\pi/a_0)$ and $k=(\pi/a_0,0)$, the wavelengths of incommensurate modulations $q_1^*$ and $q_5^*$ are controlled by its $k$-space location.[59] Equivalent phenomena have also been reported for $Bi_2Sr_2CuO_{6+\delta}$.[57] A natural speculation arising from all these observations is that scattering related to antiferromagnetic fluctuations is involved in *both* the disappearance of the Bogoliubov QPI patterns and the appearance of the incommensurate quasi-static modulations at $\mathbf{q}_1^*$ and $\mathbf{q}_5^*$ at the diagonal lines between $\mathbf{k}=(0,\pi/a_0)$ and $\mathbf{k}=(\pi/a_0,0)$ within the $CuO_2$ Brillouin zone.[103]

If the PG state of underdoped cuprates is a phase incoherent *d*-wave superconductor, these Bogoliubov-like QPI octet interference patterns could continue to exist above the transport $T_c$. This is because, if the quantum phase $\phi(\mathbf{r},t)$ is fluctuating while the energy gap magnitude $\Delta(\mathbf{k})$ remains largely unchanged, the particle-hole symmetric octet of high joint-density-of-states regions generating the QPI should continue to exist.[104,105,106] However, any gapped $\mathbf{k}$-space regions supporting Bogoliubov-like QPI in the PG phase must then occur beyond the tips of the ungapped Fermi Arc.[42] Phenomena indicative of phase fluctuating superconductivity are detectable for cuprates in particular regions of the phase diagram[107,108,109,110,111,112] as indicated by the region $T_c<T<T_\phi$ (Fig. 1(a)). The techniques involved include terahertz transport studies,[107] the Nernst effect,[108,109] torque-magnetometry measurements,[110] field dependence of the diamagnetism,[111] and zero-bias conductance enhancement.[112] Moreover, because cuprate superconductivity is quasi-two-dimensional, the superfluid density increases from zero approximately linearly with *p*, and the superconducting energy gap $\Delta(\mathbf{k})$ exhibits four $\mathbf{k}$-space nodes, fluctuations of the $\phi(\mathbf{r},t)$ of the order parameter $\Psi=\Delta(\mathbf{k})e^{i\phi(\mathbf{r},t)}$ could strongly impact the superconductivity at low *p*.[17-22]

To explore these issues, the temperature evolution of the Bogoliubov octet in $Z(\mathbf{q},E)$ was studied as a function of increasing temperature from the dSC phase into the PG phase using a 48nm square FOV. Representative $Z(\mathbf{q},E)$ for six temperatures are shown in Fig. 8; the $\mathbf{q}_i(E)$ ($i$=1,2,...,7) characteristic of the superconducting octet model are observed to remain unchanged upon passing above $T_c$ to at least $T \sim 1.5T_c$. This demonstrates that the Bogoliubov-like QPI octet phenomenology exists in the cuprate PG phase (although it is generated by different regions of $\mathbf{k}$-space, and thus different $\Delta(\mathbf{k})$, than in the same sample in the SC phase). Thus for the low-energy ($|E|<35$mV) excitations in the PG phase, the $\mathbf{q}_i(E)$ ($i$=1,2,...,7) characteristic of the octet model are preserved unchanged upon passing above $T_c$. Importantly, all seven $\mathbf{q}_i(E)$ ($i$=1,2,...,7) modulation wavevectors which are dispersive in the dSC phase remain dispersive into the PG phase still consistent with the octet model.[60] The octet wavevectors also retain their particle-hole symmetry $\mathbf{q}_i(+E) = \mathbf{q}_i(-E)$ in the PG phase and the $g(\mathbf{r},E)$ modulations occur in the same energy range and emanate from the same contour in **k**-space as those observed at lowest temperatures.[60] However, with increasing $T$ the particle-hole symmetric energy gap $\Delta(\mathbf{k})$ closes near the nodes, leaving behind a growing Fermi arc of gapless excitations.

Thus the Bogoliubov QPI signatures detectable in the dSC phase survive virtually unchanged into the underdoped PG phase - up to at least $T\sim1.5T_c$ for strongly underdoped $Bi_2Sr_2CaCu_2O_{8+\delta}$ samples. Additionally, for $|E| \leq \Delta_0$ all seven dispersive $\mathbf{q}_i(E)$ modulations characteristic of the octet model in the dSC phase remain dispersive in the PG phase. These observations rule out the existence for all $|E| \leq \Delta_0$ of non-dispersive $g(E)$ modulations at finite ordering wavevector $\mathbf{Q}^*$ which would be indicative of a static electronic order which breaks translational symmetry, a conclusion which is in agreement with the results of ARPES studies.[100, 101] In fact the excitations observable using QPI are indistinguishable from the dispersive **k**-space eigenstates of a phase incoherent d-wave superconductor.[60]

Our overall picture of electronic structure in the strongly underdoped PG phase from SI-STM contains three elements: (i) the ungapped Fermi arc,[42] (ii) the particle-hole symmetric gap $\Delta(\mathbf{k})$ of a phase incoherent superconductor,[60] and (iii) the locally symmetry breaking excitations at the $E\sim\Delta_1$ energy scale[51,58,59,60,69,70] (which remain completely unaltered upon the transition between the dSC and the PG phases[60,69]). This three-component description of the electronic structure of the cuprate pseudogap phase (Fig. 12) has recently been confirmed in detail by ARPES studies[47].

Subsequent to a detailed discussion of the $|E|\sim\Delta_1$ states in § 7 and § 8, all these QPI phenomena are summarized in context in § 9.

## 7.    Broken Spatial Symmetries of $E\sim\Delta_1$ Pseudogap States

In general for underdoped cuprates, the electronic excitations in the pseudogap energy range $|E|\sim\Delta_1$ are observed to be highly anomalous. They are associated with a strong antinodal pseudogap in **k**-space,[9,10] they exhibit slow dynamics without recombination to form Cooper pairs,[38] their Raman characteristics appear distinct from expectations for a d-wave superconductor,[40] and they appear not to contribute to superfluid density,[41] As described in § 4, § 5 and especially § 6, underdoped cuprates exhibit an octet of dispersive Bogoliubov QPI wavevectors $\mathbf{q}_i(E)$, but only upon a limited and doping-dependent arc in **k**-space. Surrounding the pseudogap energy $E\sim\Delta_1$, these phenomena are replaced by a spectrum of states whose dispersion is extremely slow (Fig. 6(c)).[51,57,58,59, 60, 69] Atomically resolved **r**-space images of the phenomena in $Z(\mathbf{r}, E \sim \Delta_1)$ show highly similar spatial patterns but with variations of intensity due to the $\Delta_1$-disorder (Fig. 5(a)). By changing to reduced energy variables $e(\mathbf{r}) = E / \Delta_1(\mathbf{r})$ and imaging $Z(\mathbf{r}, e)$ it becomes clearer that these modulations exhibit a strong maximum in intensity at $e = 1$[59,60] and that they locally break translational symmetry, and reduce the expected $C_4$ symmetry of states within the unit cell to at least to $C_2$ symmetry.[58, 59, 60,69,70]

Theoretical concerns [113] about a possibly spurious nature to spatial symmetry breaking in these $Z(\mathbf{r}, E \sim \Delta_1)$ images were addressed by carrying out a sequence of identical experiments on two very different cuprates: strongly underdoped $Ca_{1.88}Na_{0.12}CuO_2Cl_2$ ($T_c \sim$ 21 K) and $Bi_2Sr_2Ca_{0.8}Dy_{0.2}Cu_2O_{8+\delta}$ ($T_c \sim$ 45 K). These materials have completely different crystallographic structures, chemical constituents, dopant-ion species, and inequivalent dopant sites within the crystal-termination layers,[58] but were studied at the same hole-density $p$=10%. Images of the $|E| \sim \Delta_1$ states for these two systems demonstrate statically indistinguishable electronic structure arrangements.[58] As these virtually identical phenomena at $|E| \sim \Delta_1$ in these two materials must occur due to the common characteristic of these two quite different materials, the spatial characteristics of $Z(\mathbf{r}, e$=1) images[58,59,60,69] should be ascribed to the intrinsic electronic structure of the $CuO_2$ plane.

To explore the broken spatial symmetries of the $|E| \sim \Delta_1$ states within the $CuO_2$ unit cell, we used high-resolution $Z(\mathbf{r}, e)$ imaging performed on multiple different underdoped $Bi_2Sr_2CaCu_2O_{8+\delta}$ samples with $T_c$'s between 20K and 55K The necessary registry of the Cu sites in each $Z(\mathbf{r}, e)$ is achieved by the picometer scale transformation that renders the topographic image $T(\mathbf{r})$ perfectly $a_0$-periodic (§ 3). The same transformation is then applied to the simultaneously acquired $Z(\mathbf{r}, e)$ to register all the electronic structure data to this ideal lattice. The topograph $T(\mathbf{r})$ is shown in Fig. 9(a); the inset compares the Bragg peaks of its real (in-phase) Fourier components Re $T(Q_x)$, Re $T(Q_y)$ showing that Re$T(Q_x)$ / Re$T(Q_y) = 1$. Therefore $T(\mathbf{r})$ preserves the $C_4$ symmetry of the crystal lattice. In contrast, Fig. 9(b) shows that the $Z(\mathbf{r}, e = 1)$ determined simultaneously with Fig. 9(a) breaks various crystal symmetries.[58-60] The inset shows that since Re$Z(Q_x, e = 1)/$Re$Z(Q_y, e = 1) \neq 1$ the pseudogap states break $C_4$ symmetry. We therefore defined a normalized measure of intra-unit cell nematicity as a function of $e$:

$$O_N^Q(e) \equiv \frac{\text{Re}\,Z(\mathbf{Q}_y, e) - \text{Re}\,Z(\mathbf{Q}_x, e)}{\overline{Z}(e)} \qquad (11)$$

where $\overline{Z}(e)$ is the spatial average of $Z(\mathbf{r}, e)$. The plot of $O_N^Q(e)$ in Fig. 9(c) shows that the magnitude of $O_N^Q(e)$ is low for $e << \Delta_0/\Delta_1$, begins to grow near $e \sim \Delta_0/\Delta_1$, and becomes well defined as $e \sim 1$ or $E \sim \Delta_1$. Within the $CuO_2$ unit cell itself we directly imaged $Z(\mathbf{r}, e)$ [58,69] to explore where the symmetry breaking stems from. Figure 9(d) shows the topographic image of a representative region from Fig. 9(a); the locations of each Cu site $\mathbf{R}$ and of the two O atoms within its unit cell are indicated. Figure 9(e) shows $Z(\mathbf{r}, e)$ measured simultaneously with Fig. 9(d) with same Cu and O site labels. An $\mathbf{r}$-space measure of intra-unit-cell nematicity in can then be defined

$$O_N^R(e) = \sum_{\mathbf{R}} \frac{Z_x(\mathbf{R}, e) - Z_y(\mathbf{R}, e)}{\overline{Z}(e) N} \qquad (12)$$

where $Z_x(\mathbf{R}, e)$ is the magnitude of $Z(\mathbf{r}, e)$ at the O site $a_0/2$ along the $x$-axis from $\mathbf{R}$ while $Z_y(\mathbf{R}, e)$ is the equivalent along the $y$-axis, and $N$ is the number of unit cells. This estimates intra-unit-cell nematicity similarly to $O_N^Q(e)$ but counting only O site contributions. Fig. 9(f) contains the calculated value of $O_N^R(e)$ from the same FOV as Figs. 9(a) and 9(b) showing the good agreement with $O_N^Q(e)$.

The incommensurate modulations of the $|E| \sim \Delta_1$ states exhibit two ultra-slow dispersion $\mathbf{q}$-vectors, $\mathbf{q_1}^*$ and $\mathbf{q_5}^*$, which evolve with $p$ as shown in Figs. 7(b) and 7(c). The $\mathbf{q_1}^*$ modulations appear as the energy transitions from below to above $\Delta_0$ but disappear quickly leaving only two primary electronic structure elements of the pseudogap-energy electronic structure in $Z(\mathbf{q}, E \cong \Delta_1)$. These occur at $\mathbf{Q}_x = (1,0)2\pi/a_0$ and $\mathbf{Q}_y = (0,1)\,2\pi/a_0$ (the Bragg peaks representing the periodicity of the $CuO_2$ unit cell) and at the incommensurate wavevector $\mathbf{S_x}$, $\mathbf{S_y}$ which locally break translational and rotational symmetry at the nanoscale. The doping evolution of $|\mathbf{S_x}| = |\mathbf{S_y}|$ (which is by definition that of $\mathbf{q_5}^*$ - see Figs. 7(b) and 7(c)) indicates that these modulations are directly and fundamentally linked to the doping-dependence of the extinction point of the arc of Bogoliubov QPI.

The rotational symmetry breaking of these incommensurate smectic modulations can be examined by defining a measure analogous to eq. (11) of $C_4$ symmetry breaking, but now focused only upon the modulations with $\mathbf{S_x}$, $\mathbf{S_y}$ :

$$O_S^Q(e) = \frac{\operatorname{Re} Z(\mathbf{S_y}, e) - \operatorname{Re} Z(\mathbf{S_x}, e)}{\overline{Z}(e)} \qquad (13)$$

Low values found for $\left|O_S^Q(e)\right|$ at low $e$ occur because these states are dispersive Bogoliubov quasiparticles[54,55,57,60)] and cannot be analysed in term of any static electronic structure, smectic or otherwise but $\left|O_S^Q(e)\right|$ shows no tendency to become well established at the pseudogap or any other energy.[69)]

In summary: $Z(\mathbf{r}, E)$ images in underdoped $Bi_2Sr_2CaCu_2O_{8+\delta}$ reveal compelling evidence for intra-unit-cell $C_4$ symmetry breaking specific to the states at the $|E| \sim \Delta_1$ pseudogap energy. This highly disordered intra-unit-cell nematicity coexists with finite $\mathbf{Q} = \mathbf{S_x}$, $\mathbf{S_y}$ smectic electronic modulations, but they can be analyzed separately by using Fourier filtration techniques. These two types of electronic phenomena represent clearly distinct broken symmetries. The wavevector of smectic electronic modulations is controlled by the point in $\mathbf{k}$-space where the Bogoliubov interference signature disappears when the arc supporting delocalized Cooper pairing approaches the lines between $\mathbf{k} = \pm(0, \pi/a_0)$ and $\mathbf{k} = \pm(\pi/a_0, 0)$ (see Figs. 6(b) and 6(c)). This appears to indicate that the $\mathbf{Q} = \mathbf{S_x}$, $\mathbf{S_y}$ smectic effects are dominated by the same $\mathbf{k}$-space phenomena which restrict the regions of Cooper pairing[59)] and are not a characteristic of $\mathbf{r}$-space ordering.

## 8.    Interplay of Intra-unit-cell and Incommensurate Broken Symmetry States

The distinct properties of the $|E| \sim \Delta_1$ smectic modulations can be examined independently of the $|E| \sim \Delta_1$ intra-unit-cell $C_4$-symmetry breaking, by focusing in $\mathbf{q}$-space only upon the incommensurate modulation peaks $\mathbf{S_x}$ and $\mathbf{S_y}$. A coarse grained image of the local smectic symmetry breaking reveals the very short correlation length of the strongly disordered smectic modulations.[58,69,70)] The amplitude and phase of two unidirectional incommensurate modulation components measured in each $Z(\mathbf{r}, e = 1)$ image (Figs. 10(a) and 10(b)) can be further extracted by denoting the local contribution to the $\mathbf{S_x}$ modulations at position $\mathbf{r}$ by a complex field $\Psi_1(\mathbf{r})$. This contributes to the $Z(\mathbf{r}, e = 1)$ data as $\Psi_1(\mathbf{r})e^{i\mathbf{S_x} \cdot \mathbf{r}} + \Psi_1^*(\mathbf{r})e^{-i\mathbf{S_x} \cdot \mathbf{r}} \equiv 2\left|\Psi_1(\mathbf{r})\right|\cos(\mathbf{S_x} \cdot \mathbf{r} + \phi_1(\mathbf{r}))$ thus allowing the local phase $\phi_1(\mathbf{r})$ of $\mathbf{S_x}$ modulations to be mapped; similarly for the local phase $\phi_2(\mathbf{r})$ of $\mathbf{S_y}$ modulations.

A typical example of an individual topological defect (within solid white box in Fig. 10(a)) is shown in Figs. 10(c) and 10(d). The dislocation core (Fig. 10(c)) and its associated $2\pi$ phase winding (Fig. 10(d)) are clear. We find that the amplitude of $\Psi_1(\mathbf{r})$ or $\Psi_2(\mathbf{r})$ always goes to zero near each topological defect. In Figs. 10(e) and 10(f) we show the large FOV images of $\phi_1(\mathbf{r})$ and $\phi_2(\mathbf{r})$ derived from $Z(\mathbf{r}, e =1)$ in Figs. 10(a) and 10(b). They show that the smectic phases $\phi_1(\mathbf{r})$ and $\phi_2(\mathbf{r})$ take on all values between 0 and $\pm 2\pi$ in a complex spatial pattern. Large numbers of topological defects with $2\pi$ phase winding are observed; these are indicated by black ($+2\pi$) and white ($-2\pi$) circles and occur in approximately equal numbers. These data are all in agreement with the theoretical expectations for quantum smectic dislocations.[33]

Simultaneous imaging of two different broken symmetries in the electronic structure provides an unusual opportunity to explore their relationship empirically, and to develop a Ginsburg-Landau style description of their interactions. The local nematic fluctuation $\delta O_n(\vec{r}) \equiv O_n(\vec{r}) - \langle O_n \rangle$ (Fig. 11(a)) is the natural small quantity to enter the GL functional. Because, while local intra-unit-cell $C_4$ breaking is observed universally, $\langle O_n \rangle = 0$ is often found at higher dopings, the expansion should be in terms of $O_n(\vec{r})$ itself in that case. In all cases we then focus upon the phase fluctuations of the smectic modulations, meaning that $\delta O_n(\vec{r})$ couples to local shifts of the wavevectors $\vec{S}_x$ and $\vec{S}_y$. Replacing the gradient in the x- direction by a covariant-derivative-like coupling:

$$\nabla_x \psi_1(\vec{r}) \rightarrow (\nabla_x + i c_x \delta O_n(\vec{r})) \psi_1(\vec{r}), \tag{14}$$

and similarly for the y-direction, yields a GL term coupling the nematic to smectic states. The vector $\vec{c} = (c_x, c_y)$ represents by how much the wavevector $\vec{S}_x$, is shifted for a given fluctuation $\delta O_n(\vec{r})$. Hence, we proposed a GL functional (for $\vec{S}_x$) based on symmetry principles and $\delta O_n(\vec{r})$ and $\psi_1(\vec{r})$ being small:

$$F_{GL}[\delta O_n, \psi_1] = F_n[\delta O_n] + \int d^2 r \left( a_x |(\nabla_x + i c_x \delta O_n) \psi_1|^2 + a_y |(\nabla_y + i c_y \delta O_n) \psi_1|^2 + m |\psi_1|^2 + ... \right) \tag{15}$$

where … refers to terms we can neglect. It is interesting to note that if one replaced $\vec{c} \delta O_n(\vec{r})$ by $\frac{2e}{\hbar} \vec{A}(\vec{r})$ where $\vec{A}(\vec{r})$ is the electromagnetic vector potential, Equation (15) would become the familiar GL free energy of a superconductor. In that well known case, minimization yields $\vec{A}(\vec{r}) = \frac{\hbar}{2e}\vec{\nabla}\varphi(\vec{r})$ and thus quantization of magnetic flux.[114] Analogously, minimization of eq. (15) implies $\delta\mathcal{O}_n(\vec{r}) = \vec{l}\cdot\vec{\nabla}\varphi$ surrounding each topological defect[70] where the vector $\vec{l} \propto (\alpha_x, \alpha_y)$ and lies along the line where $\delta\mathcal{O}_n(\vec{r}) = 0$. The resulting prediction is that $\delta\mathcal{O}_n(\vec{r})$ will vanish along the line in the direction of $\vec{l}$ that passes through the core of the topological defect with $\mathcal{O}_n(\vec{r})$ becoming greater on one side and less on the other (Fig. 11(b)).

To demonstrate that this GL functional captures the observed $\delta\mathcal{O}_n - \psi_s$ coupling in $Bi_2Sr_2CaCu_2O_{8+\delta}$(Fig. 11(a)), we apply eq. (15) including both $\vec{S}_x$ and $\vec{S}_y$ smectic modulations and simulate the profile of $\delta\mathcal{O}_n(\vec{r})$ treating the phase and amplitude of smectic fields $\psi_1(\vec{r})$ and $\psi_2(\vec{r})$ as mean-field input to determine $\delta\mathcal{O}_n(\vec{r})$. Figures 11(c) and 11(d) show the overlay of topological defect locations within the boxes in Fig. 11(a) on $\delta\mathcal{O}_n(\vec{r})$ as simulated using eq. (15). This demonstrates directly how the GL functional associates fluctuations in $\delta\mathcal{O}_n(\vec{r})$ with the smectic topological defect locations in the fashion of Fig. 11(b). The close similarity between the measured $\delta\mathcal{O}_n(\vec{r})$ in Figs. 11(e) and 11(f) and the simulation in Figs. 11(c) and 11(d) demonstrates how the minimal GL model of eq. (15) captures the interplay between the measured intra-unit-cell nematicity fluctuations $\delta\mathcal{O}_n(\vec{r})$ (Fig. 11(a)) and disordered smectic modulations (Fig. 10). These G-L parameters vary somewhat from location to location due to extrinsic disorder.[70]

Both nematic and smectic broken symmetries have been reported in the electronic/magnetic structure of different cuprate compounds.[115,116,117,118] If the tendency for intra-unit-cell nematicity and disordered smectic modulations to coexist[69,70] summarized here is ubiquitous to underdoped cuprates, which broken symmetry manifests at the macroscopic scale will depend on the coefficients in the GL functional given in eq. (15) and on other material specific aspects such as crystal symmetry. This approach may provide a good starting point to address the interplay between the different broken electronic symmetries and the superconductivity near the Mott insulator state.

## 9. Conclusions and Future

A fundamentally bipartite electronic structure in strongly underdoped cuprates approaching the Mott insulator emerges from SI-STM studies as summarized by Fig. 12. In the dSC phase (Figs. 12(a), 12(b) and 12(c)) the Bogoliubov QPI signature of delocalized Cooper pairs (§ 6) exists upon the arc in **k**-space labeled by region II in Fig. 12(b). These states have energy $|E| \le \Delta_0$. The Bogoliubov QPI disappears near the lines connecting **k**=$(0,\pm\pi/a_0)$ to **k**=$(\pm\pi/a_0,0)$ - thus defining a **k**-space arc which supports the delocalized Cooper pairing. This arc shrinks rapidly towards the **k**=$(\pm\pi/2a_0,\pm\pi/2a_0)$ points with falling hole-density in a fashion which could satisfy Luttinger's theorem if it were actually a hole-pocket bounded from behind by the **k**=$\pm(\pi/a_0,0)$ - **k**=$\pm(0,\pi/a_0)$ lines. The $|E|\sim\Delta_1$ pseudogap excitations (§ 7) are labeled schematically by region I in Fig. 12(b). They exhibit a radically different **r**-space phenomenology locally breaking the expected $C_4$ symmetry of electronic structure at least down to $C_2$ and possibly to an even lower symmetry, within each $CuO_2$ unit cell (Fig. 12(a)). These intra-unit-cell broken $C_4$-symmetry states coexist with incommensurate modulations that break translational and rotational symmetry locally. The wavelengths of these incommensurate modulations **Q**=$\mathbf{S_x}$, $\mathbf{S_y}$ are controlled by the **k**-space locations at which the Bogoliubov QPI signatures disappear; this is the empirical reason why $\mathbf{S_x}$, $\mathbf{S_y}$ evolve continuously with doping along the line joining **k**=$(0,\pm\pi/a_0)$ - **k**=$(\pm\pi/a_0,0)$ (§ 6). In the PG phase (Figs. 12(d), 12(e) and 12(f)), the Bogoliubov QPI signature (Fig. 12(f)) exists upon a smaller part of the same arc in **k**-space as it did in the dSC phase. This is labeled as region II in Fig. 12(e). Here, however, since the ungapped Fermi-arc (region III) predominates, the gapped region supporting $d$-wave Bogoliubov QPI has shrunk into a narrow sliver inside the line connecting **k**=$(\pi/a_0,0)$ and **k**=$(0,\pi/a_0,)$ (Fig. 12(e)). The $E\sim\Delta_1$ excitations in the PG phase, (§ 7) are again labeled by region I and exhibit intra-unit-cell $C_4$ breaking and **Q**=$\mathbf{S_x}$, $\mathbf{S_y}$ incommensurate smectic modulations indistinguishable from those in the dSC phase (Fig. 12(e)).

The relationship between the $|E|\sim\Delta_1$ broken symmetry states (§ 5, § 7 and § 8) and the $|E|\le\Delta_0$ Bogoliubov quasiparticles indicative of Cooper pairing (§ 4 and § 6) is not yet understood. However, these two sets of phenomena appear to be linked inextricably. The reason is that the **k**-space location where the latter disappears always occurs where the Fermi surface touches the lines connecting **k**=$(0,\pm\pi/a_0)$ to **k**=$(\pm\pi/a_0,0)$, while the wavevectors $\mathbf{q_1}^*$ and $\mathbf{q_5}^*$ close to this intersection are those of the incommensurate modulations at $|E|\sim\Delta_1$. One stimulating conjecture arising from these observations is that scattering related to antiferromagnetic fluctuations could be the cause of, and provide a natural link between, these two fundamental phenomena in the electronic structure of underdoped cuprates. One way to explore the significance of this idea for the electronic phase diagram and Cooper pairing mechanism, would be to study a hole-density where the Fermi surface does not reach the lines connecting $k=(0,\pm\pi/a_0)$ to $k=(\pm\pi/a_0,0)$ especially in the overdoped regime.

## Acknowledgements

We acknowledge and thank all our collaborators: J.W. Alldredge, I. Firmo, M.H. Hamidian, T. Hanaguri, P. J. Hirschfeld, J.E. Hoffman, E.W. Hudson, Chung Koo Kim, Y. Kohsaka, K.M. Lang, C. Lupien, Jhinhwan Lee, Jinho Lee, V. Madhavan, K. McElroy, S. Mukhopadhyay, J. Orenstein, S.H. Pan, R. Simmonds, J. Slezak, J. Sethna, H. Takagi, C. Taylor, P. Wahl, & M. Wang. Preparation of this manuscript was supported by the Center for Emergent Superconductivity, an Energy Frontier Research Center funded by the U.S. Department of Energy, Office of Basic Energy Sciences under Award Number DE-2009-BNL-PM015.

Fig. 1. (Color online) (a) Schematic copper-oxide phase diagram. Here $T_C$ is the critical temperature circumscribing a 'dome' of superconductivity, $T_\phi$ is the maximum temperature at which superconducting phase fluctuations are detectable within the pseudogap phase, and $T^*$ is the approximate temperature at which the pseudogap phenomenology first appears. (b) The two classes of electronic excitations in cuprates. The separation between the energy scales associated with excitations of the superconducting state (dSC, denoted by $\Delta_0$) and those of the pseudogap state (PG, denoted by $\Delta_1$) increases as $p$ decreases (reproduced from ref. 8). The different symbols correspond to the use of different experimental techniques. (c) A schematic diagram of electronic structure within the 1$^{st}$ Brillouin zone of hole-doped $CuO_2$. The dashed lines joining $k=(0,\pm\pi/a_0)$ to $k=(\pm\pi/a_0,0)$ are found, empirically, to play a key role in the doping-dependence of electronic structure. The putative Fermi surface is labeled using two colors, red for the 'nodal' regions bounded by the dashed lines and blue for the 'antinodal' regions near $k=(0,\pm\pi/a_0)$ and $k=(\pm\pi/a_0,0)$.

Fig. 2. (Color online) (a) Fourier transform of the conductance ratio map $Z(\mathbf{r}, E)$ at a representative energy below $\Delta_0$ for $T_C = 45K$ $Bi_2Sr_2Ca_{0.8}Dy_{0.2}Cu_2O_{8+\delta}$, which only exhibits

the patterns characteristic of homogenous $d$-wave superconducting quasiparticle interference. (b) Evolution of the spatially averaged tunneling spectra of $Bi_2Sr_2CaCu_2O_{8+\delta}$ with diminishing $p$, here characterized by $T_C(p)$. The energies $\Delta_1(p)$ (blue dashed line) are easily detected as the pseudogap edge while the energies $\Delta_0(p)$ (red dashed line) are more subtle but can be identified by the correspondence of the "kink" energy with the extinction energy of Bogoliubov quasiparticles, following the procedures in refs. 54,60. (c) Laplacian of the conductance ratio map $Z(\mathbf{r})$ at the pseudogap energy $E = \Delta_1$, emphasizing the local symmetry breaking of these electronic states.

Fig. 3. (Color online) (a) $g(\mathbf{r}, E$=-1.5mV) showing the random bright 'crosses' which are the resonant impurity states of a $d$-wave superconductor, at each Zn impurity atom site. (b) High resolution $g(\mathbf{r}, E$=-1.5mV). The bright center of scattering resonance in (b) coincides with the position of a Bi atom. The inner bright cross is oriented with the nodes of the $d$-wave gap. The weak outer features, including the ~30Å- long "quasiparticle beams" at 45$^{\mathrm{o}}$ to the inner cross, are oriented with the gap maxima. (c) The spectrum of a usual superconducting region of the sample, where Zn scatterers are absent (dark region in (a) and (b)), is shown as filled circles. The arrows indicate the superconducting coherence peaks that are suppressed near Zn. The data shown as open circles, with an interpolating fine solid line, are the spectrum taken exactly at the center of a bright Zn scattering site. It shows both the intense scattering resonance peak centered at $\Omega$=-1.5mV, and the very strong suppression of both the superconducting coherence peaks and gap magnitude at the Zn site.

Fig. 4. (Color online) (a) (b) $g(\mathbf{r}, E$=±10mV) revealing the impurity states at locations of the Ni impurity atoms in this 128 Å x128 Å square FOV. At $V_{bias}$=+10mV, showing the '+-shaped' regions of high local density of states associated with the Ni atoms. At $V_{bias}$=-10mV, showing the 45$^{\mathrm{o}}$ spatially rotated 'X-shaped' pattern. (c) $g(\mathbf{r}, E)$ spectra above a Ni atom (red) and away from the Ni atom (blue).

Fig. 5. (Color online) (a) Map of the local energy scale $\Delta_1(\mathbf{r})$ from a 49nm field of view (corresponding to ~16,000 $CuO_2$ plaquettes) measured on a sample with $T_C$= 74K. Average gap magnitude $\Delta_1$ is at the top, together with the values of $N$, the total number of dopant impurity states (shown as white circles) detected in the local spectra. (b) The average tunneling spectrum, $g(E)$, associated with each gap value in the field of view in (a). The arrows locate the "kinks" separating homogeneous from heterogeneous electronic structure and which occur at whose energy ~$\Delta_0$. (c) The doping dependence the average $\Delta_1$ (blue circles), average $\Delta_0$ (red circles) and average antinodal scattering rate $\Gamma_2$* (black squares), each set interconnected by dashed guides to the eye. The higher-scale $\Delta_1$ evolves along the pseudogap line whereas the lower-scale $\Delta_0$ represents segregation in energy between homogeneous and heterogeneous electronic structure.

Fig. 6. (Color online) (a) The 'octet' model of expected wavevectors of quasiparticle interference patterns in a superconductor with electronic band structure like that of $Bi_2Sr_2CaCu_2O_{8+\delta}$. Solid lines indicate the $\mathbf{k}$-space locations of several banana-shaped quasiparticle contours of constant energy as they increase in size with increasing energy. As an example, at a specific energy, the octet of regions of high JDOS are shown as red circles. The seven primary scattering $\mathbf{q}$-vectors interconnecting elements of the octet are shown in blue. (b) The magnitude of various measured QPI vectors, plotted as a function of energy. Whereas the expected energy dispersion of the octet vectors $q_i(E)$ is apparent for $|E| < 32$mV, the peaks which avoid extinction ($\mathbf{q_1}$* and $\mathbf{q_5}$*) ultra-slow or zero dispersion above $\Delta_0$ (vertical grey line). Inset: A plot of the superconducting energy gap $\Delta(\theta_k)$ determined from octet model inversion of quasiparticle interference measurements, shown as open circles.[56] (c) Locus of the Bogoliubov band minimum $\mathbf{k_B}(E)$ found from extracted QPI peak locations $\mathbf{q_i}(E)$, in five independent $Bi_2Sr_2CaCu_2O_{8+\delta}$ samples with decreasing hole density. Fits to quarter-circles are shown and, as $p$ decreases, these curves enclose a progressively smaller area. The BQP interference patterns disappear near the perimeter of a $\mathbf{k}$-space region bounded by the lines joining $\mathbf{k} = (0, \pm\pi/a_0)$ and $\mathbf{k} = (\pm\pi/a_0, 0)$. The spectral weights of $\mathbf{q_2}$, $\mathbf{q_3}$, $\mathbf{q_6}$ and $\mathbf{q_7}$ vanish at the same place (dashed line; see also ref. 60). Filled symbols in the inset represent the hole count $p = 1 - n$ derived using the simple Luttinger theorem, with the fits to a large, hole-like Fermi surface indicated schematically here in grey. Open symbols in the inset are the hole counts calculated using the area enclosed by the Bogoliubov arc and the lines joining $\mathbf{k} = (0, \pm\pi/a_0)$ and $\mathbf{k} = (\pm\pi/a_0, 0)$, and are indicated schematically here in blue.

Fig. 7. (Color online) (a) A large FOV $Z(\mathbf{r}, e=1)$ image from a strongly underdoped sample showing the full complexity of the electronic structure modulations. Inset: $Z(\mathbf{q}, e=1)$ for underdoped $T_C$=50K $Bi_2Sr_2CaCu_2O_{8+\delta}$. The arrows label the location of the wavevectors $\mathbf{S}_x$,

$S_y$ (or $q_5$*) and $Q_x$, $Q_y$ as described throughout the text. (b) Doping dependence of line-cuts of $Z(\mathbf{q}, E=48\text{meV})$ extracted along the Cu-O bond direction $Q_x$. The vertical dashed lines demonstrate that the $\mathbf{q}$-vectors at energies between $\Delta_0$ and $\Delta_1$ are not commensurate harmonics of a $4a_0$ periodic modulation, but instead evolve in a fashion directly related to the extinction point of the Fermi arc. (c) $q_1$*, $q_5$*, and their sum $q_1$* + $q_5$* as a function of $p$ demonstrating that, individually, these modulations evolve with doping while their sum does not change and is equal to the reciprocal lattice vector defining the first Brillouin zone. This indicates strongly that these modulations are primarily a $\mathbf{k}$-space phenomenon.

Fig. 8. (Color online) (a)-(x) Differential conductance maps $g(\mathbf{r},E)$ were obtained in an atomically resolved and registered FOV > $45 \times 45$ nm$^2$ at six temperatures. Each panel shown is the $Z(\mathbf{q},E)$ for a given energy and temperature. The QPI signals evolve dispersively with energy along the horizontal energy axis. The temperature dependence of QPI for a given energy evolves along the vertical axis. The octet-model set of QPI wave vectors is observed for every $E$ and $T$ as seen, for example, by comparing (a) and (u), each of which has the labeled octet vectors. Within the basic octet QPI phenomenology, there is no particular indication in these data of where the superconducting transition $T_c$, as determined by resistance measurements, occurs.

Fig. 9. (Color online) (a) Topographic image $T(\mathbf{r})$ of the Bi$_2$Sr$_2$CaCu$_2$O$_{8+\delta}$ surface. The inset shows that the real part of its Fourier transform Re $T(\mathbf{q})$ does not break C$_4$ symmetry at its Bragg points because plots of $T(\mathbf{q})$ show its values to be indistinguishable at $Q_x = (1, 0)2\pi/a_0$ and $Q_y = (0, 1)2\pi/a_0$. Thus neither the crystal nor the tip used to image it (and its $Z(\mathbf{r},E)$ simultaneously) exhibits C$_2$ symmetry. (b) The $Z(\mathbf{r},e=1)$ image measured simultaneously with $T(\mathbf{r})$ in (a). The inset shows that the Fourier transform $Z(\mathbf{q},e=1)$ does break C$_4$ symmetry at its Bragg points because Re $Z(Q_x, e\sim1) \neq$ Re $Z(Q_y, e\sim1)$ . (c) The value of $O_N^Q(e)$ computed from $Z(\mathbf{r}, e)$ data measured in the same FOV as (a) and (b). Its magnitude is low for all $E < \Delta_0$ and then rises rapidly to become well established near $e \sim 1$ or $E \sim \Delta_1$. Thus the pseudogap states in underdoped Bi$_2$Sr$_2$CaCu$_2$O$_{8+\delta}$ break the expected C$_4$ symmetry of CuO$_2$ electronic structure. (d) Topographic image $T(\mathbf{r})$ from the region identified by a small white box in (a). It is labeled with the locations of the Cu atom plus both the O atoms within each CuO$_2$ unit cell (labels shown in the inset). Overlaid is the location and orientation of a Cu and four surrounding O atoms. (e) The simultaneous $Z(\mathbf{r}, e = 1)$ image in the same FOV as (d) (the region identified by small white box in (b)) showing the same Cu and O site labels within each unit cell (see inset). Thus the physical locations at which the nematicity measure $O_N^g(e)$ is evaluated are labeled by the dashes. Overlaid is the location and orientation of a Cu atom and four surrounding O atoms. (f) The value of $O_N^g(e)$ computed from $Z(\mathbf{r}, e)$ data measured in the same FOV as (a) and (b). As in (c), its magnitude is low for all $E < \Delta_0$ and then rises rapidly to become well established at $e \sim 1$ or $E \sim \Delta_1$.

Fig. 10. (Color online) (a) Smectic modulations along $x$-direction are visualized by Fourier filtering out all the modulations of $Z(\vec{r}, e = 1)$ except those at $\mathbf{S_x}$, in the FOV indicated by the broken boxes in Fig. 7(a). (b) Smectic modulations along $y$-direction are visualized by Fourier filtering out all the modulations of $Z(\vec{r}, e = 1)$ except those at $\mathbf{S_y}$, in the FOV indicated by the broken boxes in Fig. 7(a). (c) Smectic modulation around the single topological defect in the same FOV showing that the dislocation core is indeed at the center of the topological defect and that the modulation amplitude tend to zero there. This is true for all the $2\pi$ topological defects identified in (e) and (f). (d) Phase field around the single topological defect in the FOV in (c). (e)((f)) Phase field $\phi_1(\mathbf{r})(\phi_2(\mathbf{r}))$ for smectic modulations along $x(y)$ - direction exhibiting the topological defects at the points around which the phase winds from 0 to $2\pi$. Depending on the sign of phase winding, the topological defects are marked by either white or black dots. The broken red circle is the measure of the spatial resolution determined by the cut-off length ($3\sigma$) in extracting the smectic field from $Z(\vec{q}, e = 1)$.

Fig. 11. (Color online) (a) Fluctuations of electronic nematicity $\delta O_n(\vec{r}, e = 1)$ obtained by subtracting the spatial average $\langle O_n(\vec{r}, e = 1) \rangle$ from $O_n(\vec{r}, e = 1)$. The locations of all $2\pi$ topological defects measured simultaneously are indicated by black dots. They occur primarily near the lines where $\delta O_n(\vec{r}, e = 1) = 0$. Inset shows the distribution of distances between the nearest $\delta O_n(\vec{r}, e = 1) = 0$ contour and each topological defect; it reveals a strong tendency for that distance to be far smaller than expected at random. The boxes show regions that are expanded in (e) and (f) and compared to simulations in (c) and (d). (b) Theoretical $\delta O_n(\vec{r}, e = 1)$ from the Ginzburg Landau functional eq. (15) at the site of a single topological defect (bottom). The vector $\vec{l}$ lies along the zero-fluctuation line of $O_n(\vec{r}, e = 1)$. (c), (d),

$\delta O_n(\vec{r}, e=1)$ obtained by numerical simulation using eq. (15) plus the experimentally obtained topological defect configurations (black dots). Red broken circle is the measure of the spatial resolution determined by the cut-off length ($3\sigma$) in extracting the smectic field. (e), (f), Measured $\delta O_n(\vec{r}, e=1)$ in the fields of view of (c), (d).

Fig. 12. (Color online) (a) Image typical of the broken spatial symmetries in electronic structure as measured in the dSC phase at the pseudogap energy $E\sim\Delta_1$ in underdoped cuprates (both $Bi_2Sr_2CaCu_2O_{8+\delta}$ and $Ca_{2-x}Na_xCuO_2Cl_2$). (b) A schematic representation of the electronic structure in one quarter of the Brillouin zone at lowest temperatures in the dSC phase. The region marked II in front of the line joining $\mathbf{k}=(\pi/a_0,0)$ and $\mathbf{k}=(0,\pi/a_0)$ is the locus of the Bogoliubov QPI signature of delocalized Cooper pairs. (c) An example of the characteristic Bogoliubov QPI signature of sixteen pairs of interference wavevectors, all dispersive and internally consistent with the octet model as well as particle-hole symmetric $\mathbf{q}_i(+E)=\mathbf{q}_i(-E)$, here measured at lowest temperatures. (d) An example of the broken spatial symmetries which are concentrated upon pseudogap energy $E\sim\Delta_1$ as measured in the PG phase; they are indistinguishable from measurements at $T\sim0$. (e) A schematic representation of the electronic structure in one quarter of the Brillouin zone at $T\sim1.5\ T_c$ in the PG phase. The region marked III is the Fermi arc, which is seen in QPI studies as a set of interference wavevectors $\mathbf{q}_i(E=0)$ which indicate that there is no gap-node at $E=0$. Region II in front of the line joining $\mathbf{k}=(\pi/a_0,0)$ and $\mathbf{k}=(0,\pi/a_0)$ is the locus of the phase incoherent Bogoliubov QPI signature. Here all 16 pairs of wavevectors of the octet model are detected and found to be dispersive. Thus although the sample is not a long-range phase coherent superconductor, it does give clear QPI signatures of $d$-wave Cooper pairs. (f) An example of the characteristic Bogoliubov QPI signature of sixteen pairs of interference wavevectors, all dispersive and internally consistent with the octet model as well as particle-hole symmetric $\mathbf{q}_i(+E)=\mathbf{q}_i(-E)$, but here measured at $T\sim1.5T_c$.

<u>References</u>

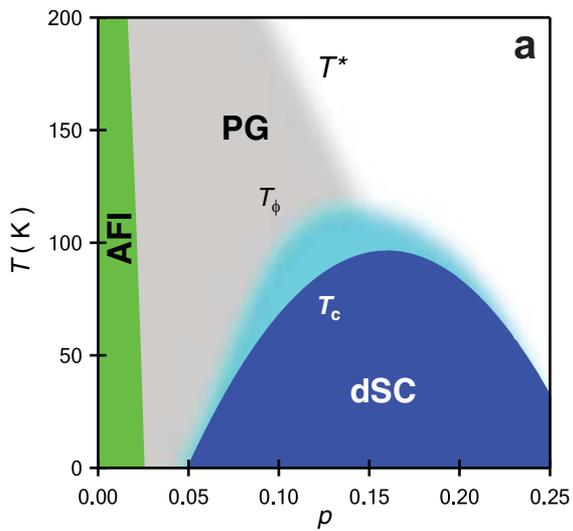

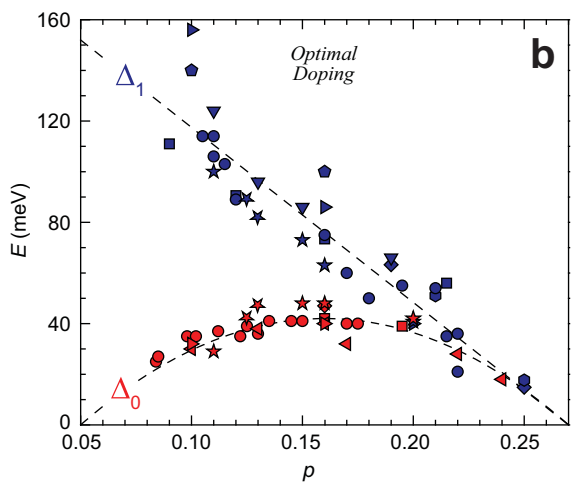

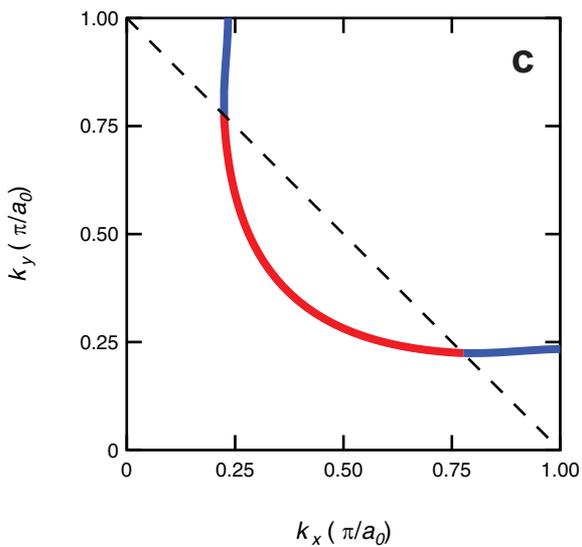

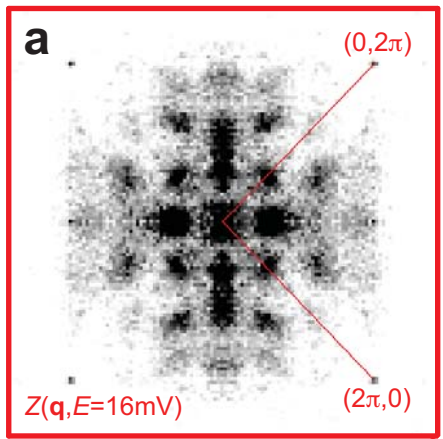

$Z(\mathbf{q}, E=16\text{mV})$

**a** $(0,2\pi)$

$(2\pi,0)$

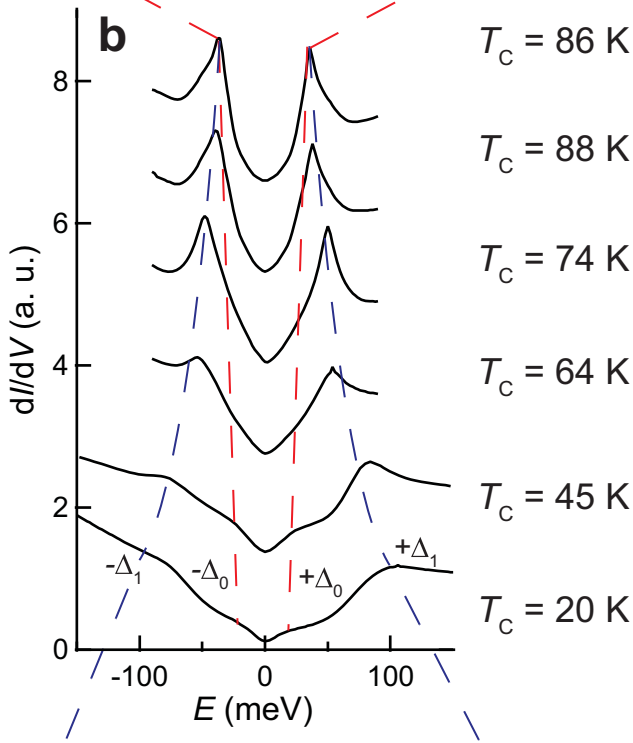

**b**

$T_C = 86$ K

$T_C = 88$ K

$T_C = 74$ K

$T_C = 64$ K

$T_C = 45$ K

$T_C = 20$ K

$-\Delta_1$ $-\Delta_0$ $+\Delta_0$ $+\Delta_1$

$dI/dV$ (a. u.)

$E$ (meV)

$-100$ $0$ $100$

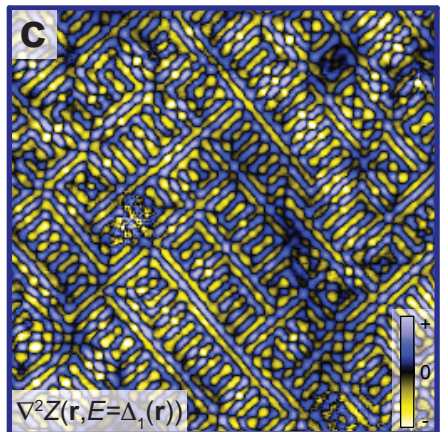

**c**

$\nabla^2 Z(\mathbf{r}, E=\Delta_1(\mathbf{r}))$

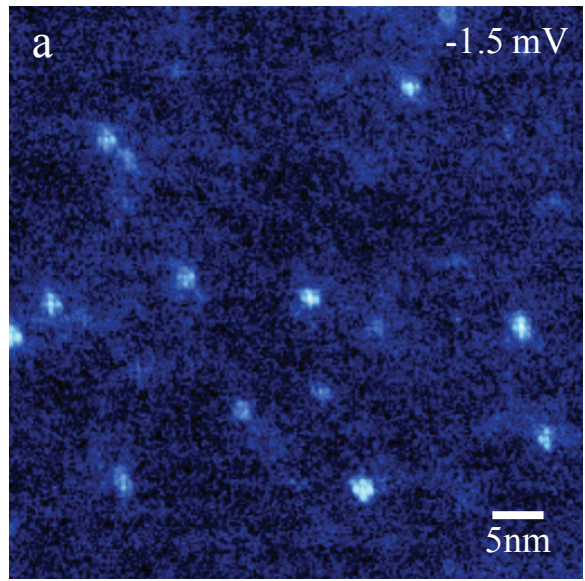

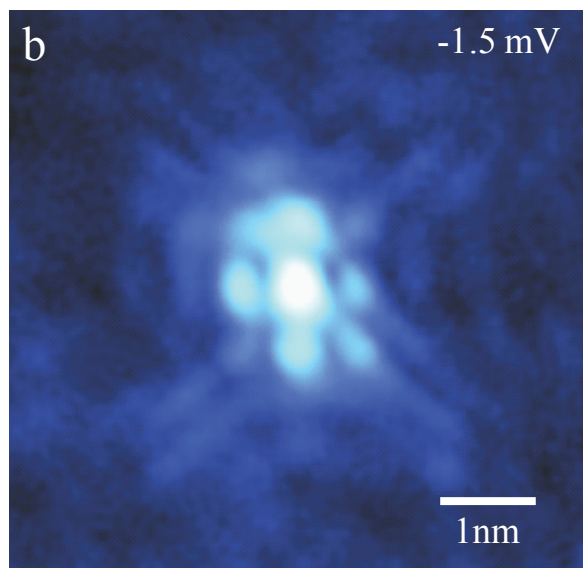

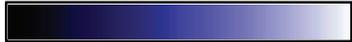

Low                    High

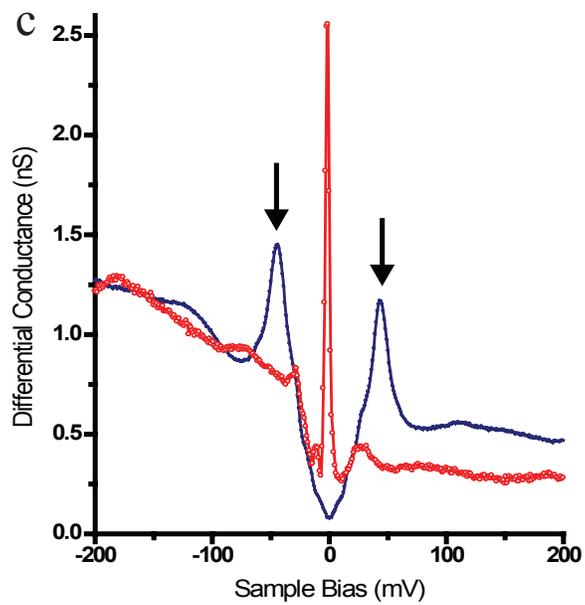

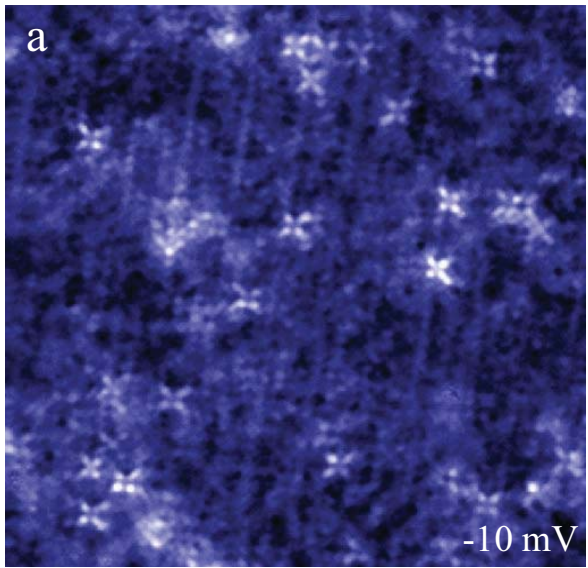

a

-10 mV

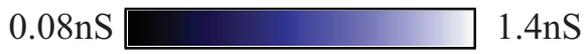

0.08nS      1.4nS

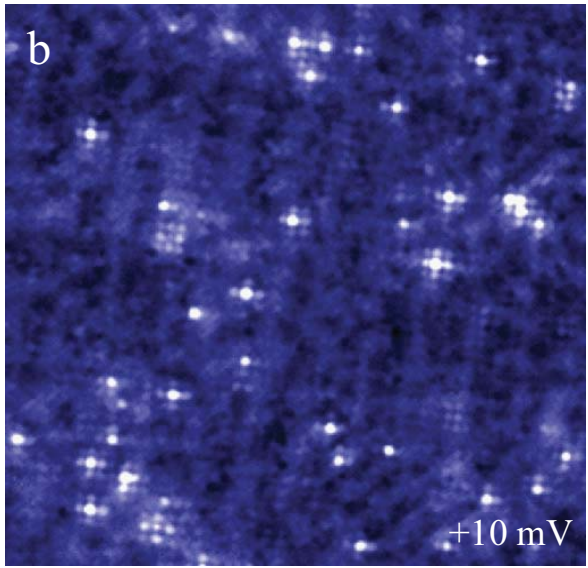

b

+10 mV

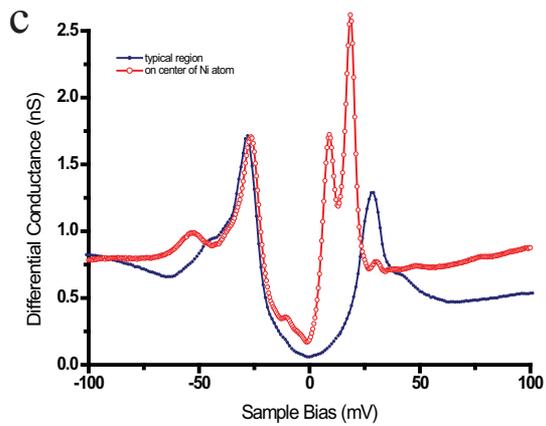

c

typical region
on center of Ni atom

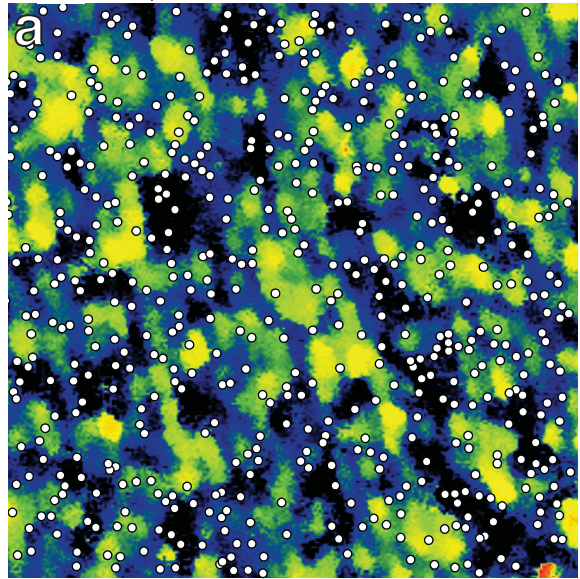

$\overline{\Delta}_1 = 55$ meV    N = 455

20 meV ▬▬▬▬▬▬ 70 meV

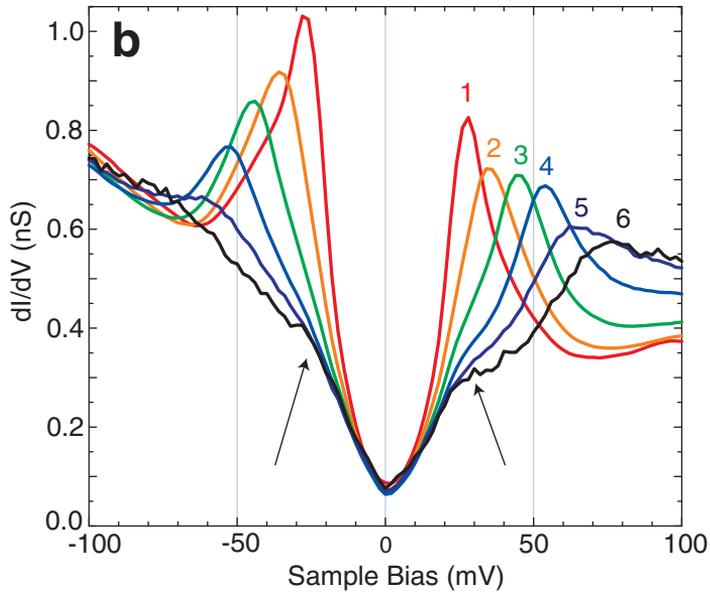

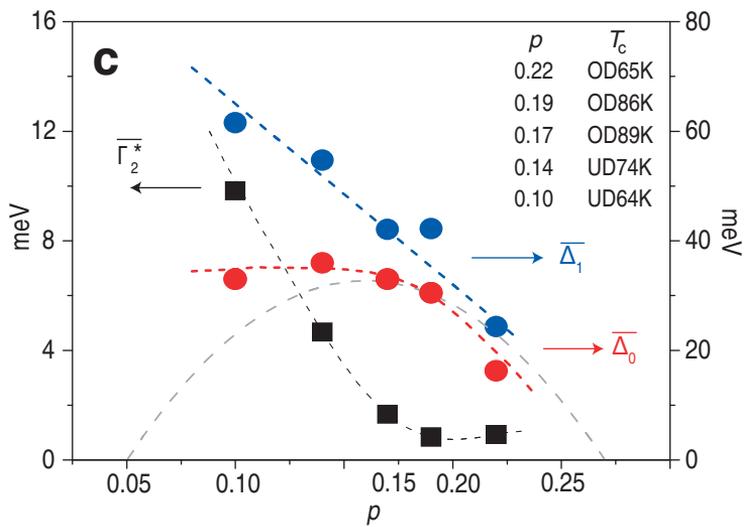

$\overline{\Gamma}_2^*$

$\overline{\Delta}_1$

$\overline{\Delta}_0$

| $p$ | $T_c$ |
|---|---|
| 0.22 | OD65K |
| 0.19 | OD86K |
| 0.17 | OD89K |
| 0.14 | UD74K |
| 0.10 | UD64K |

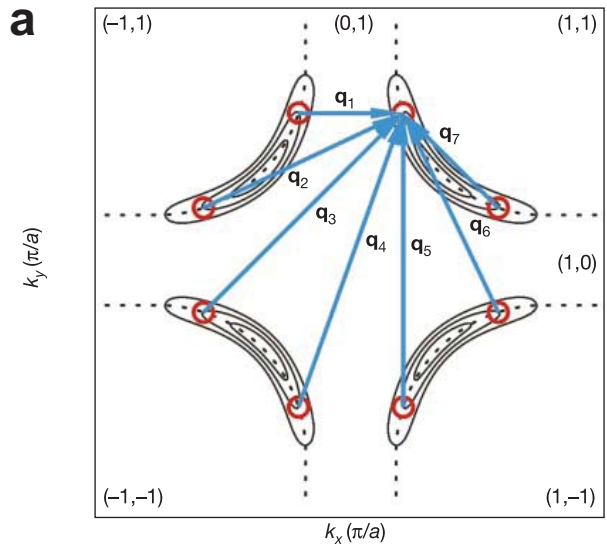

**a**

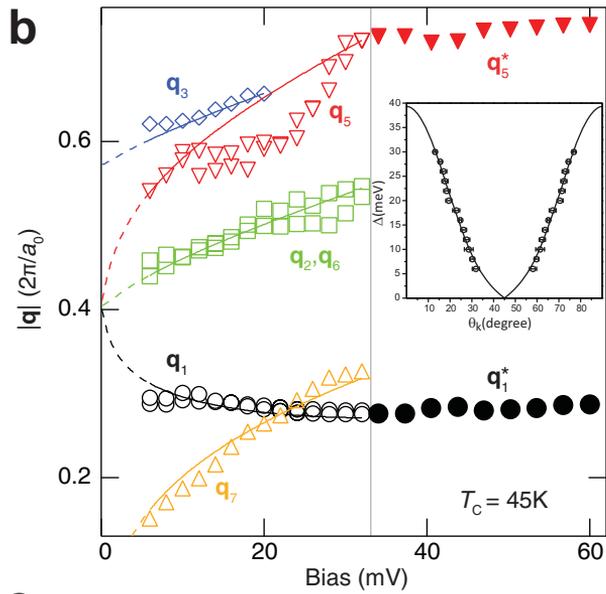

**b**

$T_c = 45K$

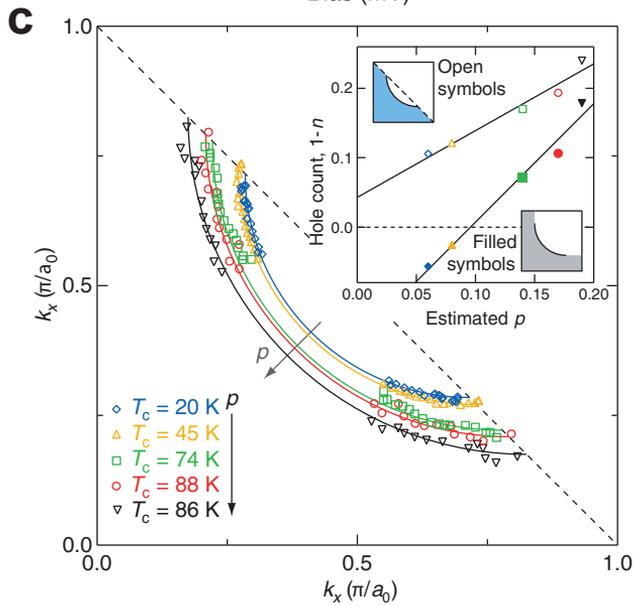

**c**

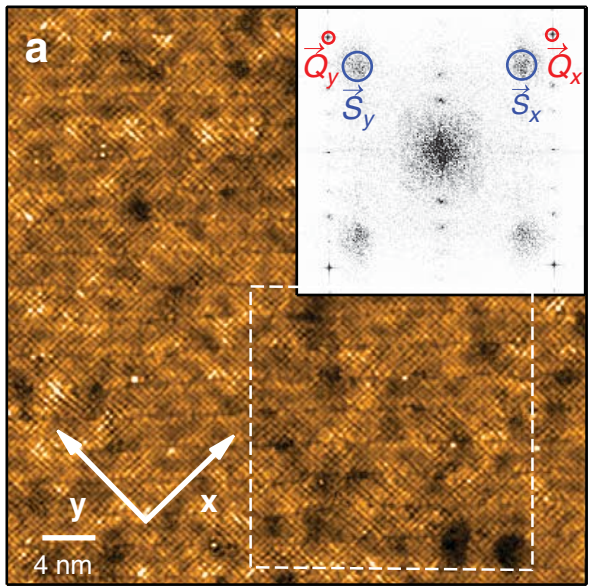

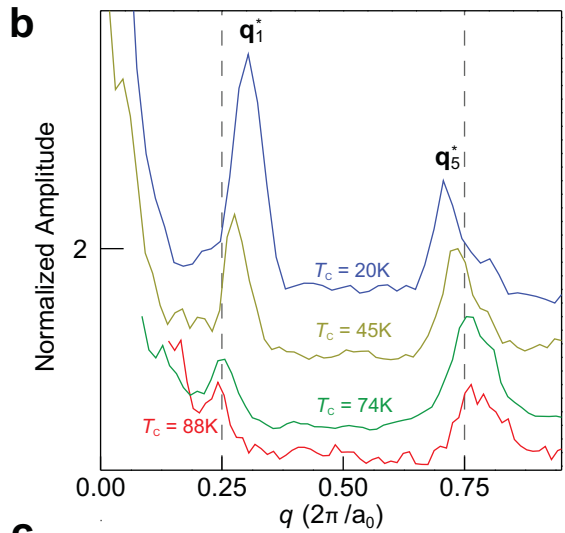

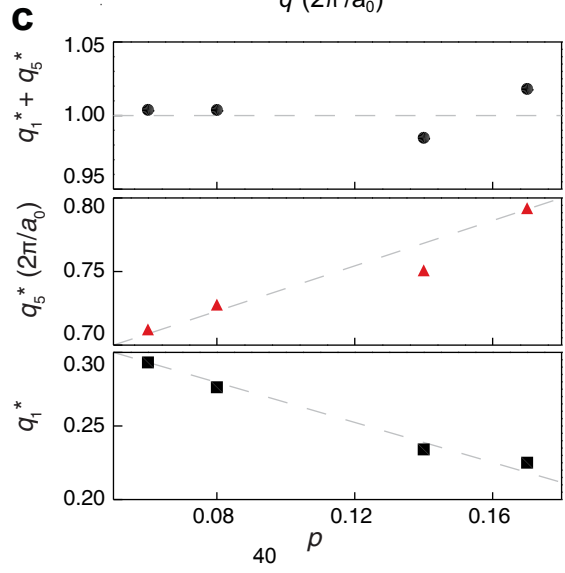

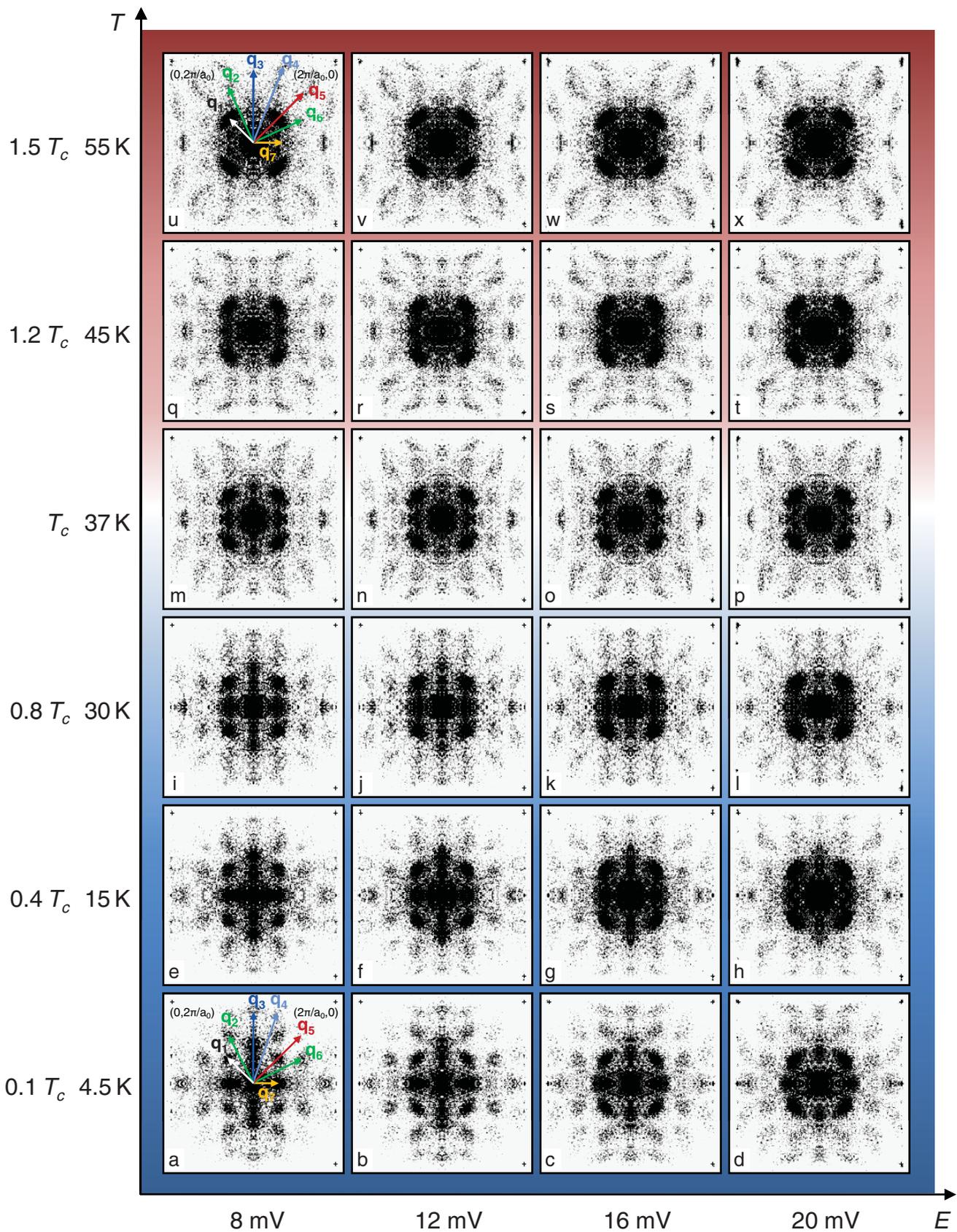

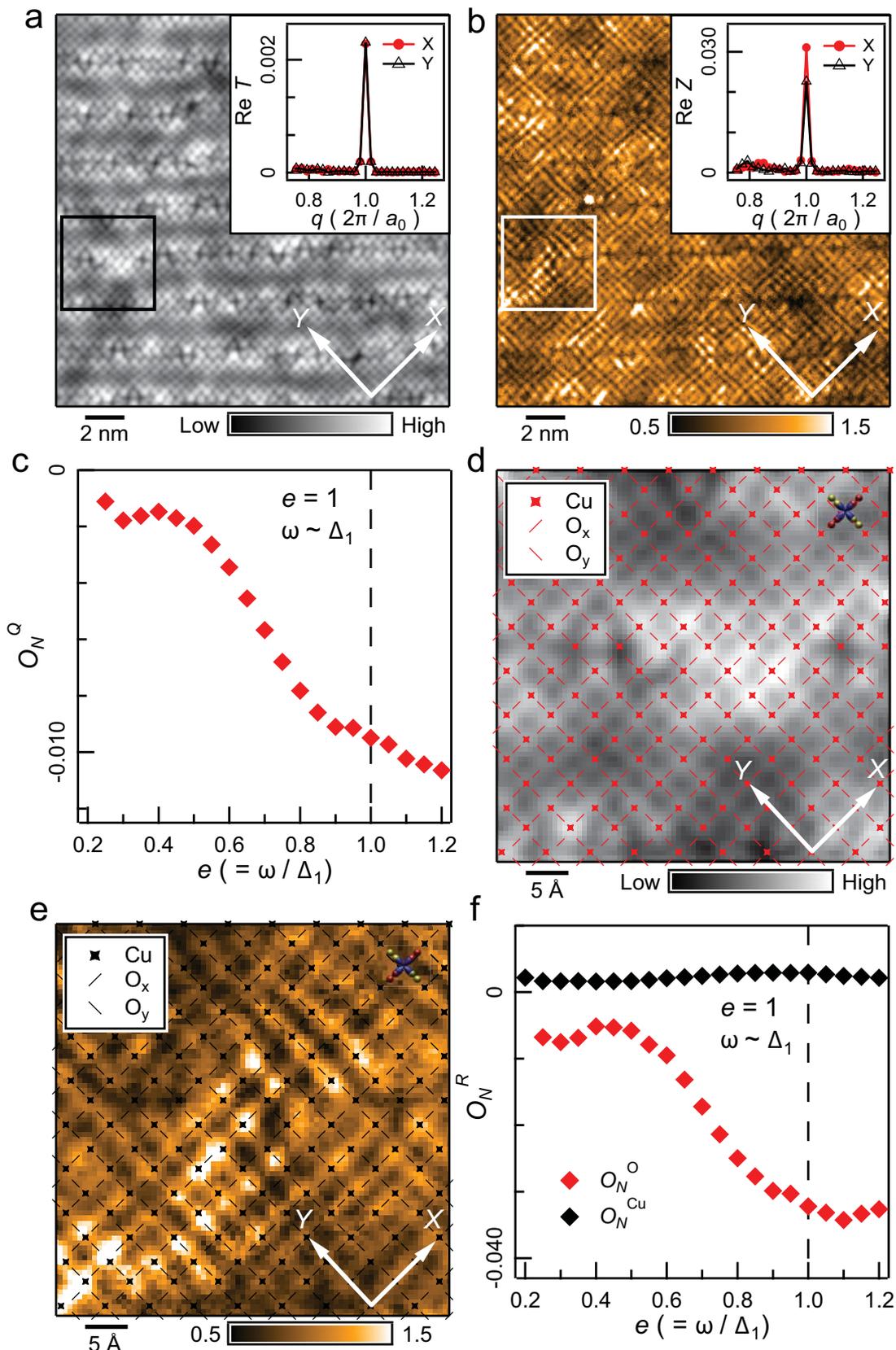

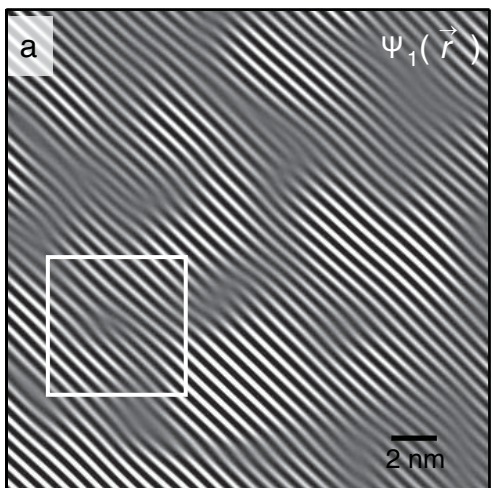
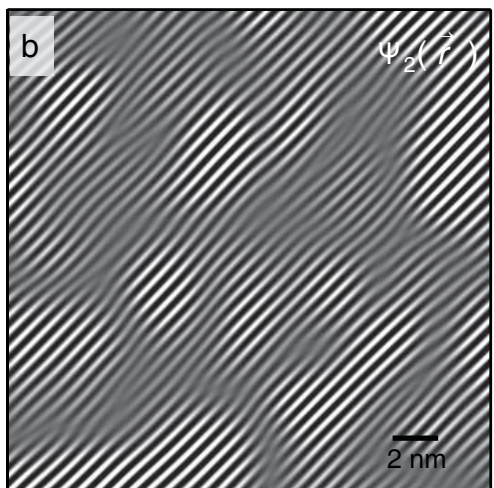

$\Psi_1(\vec{r})$

$\Psi_2(\vec{r})$

2 nm

2 nm

low ▣ high

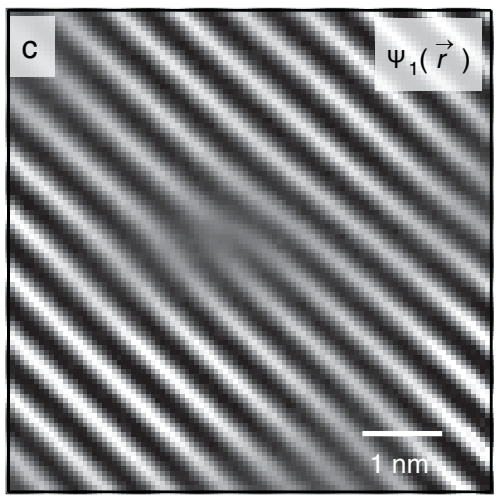
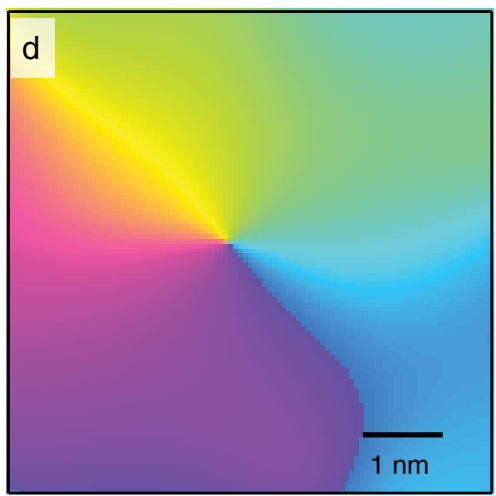

$\Psi_1(\vec{r})$

1 nm

1 nm

low ▣ high

0 ▣ $2\pi$

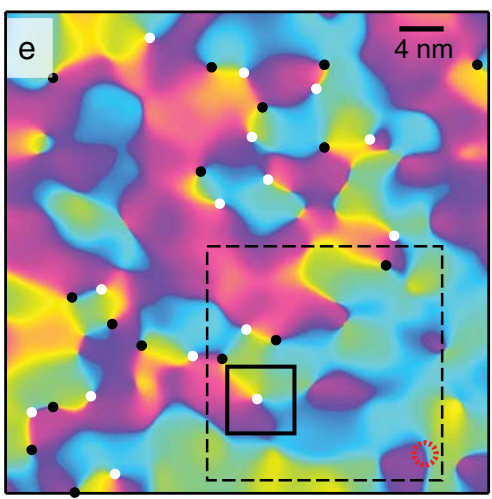
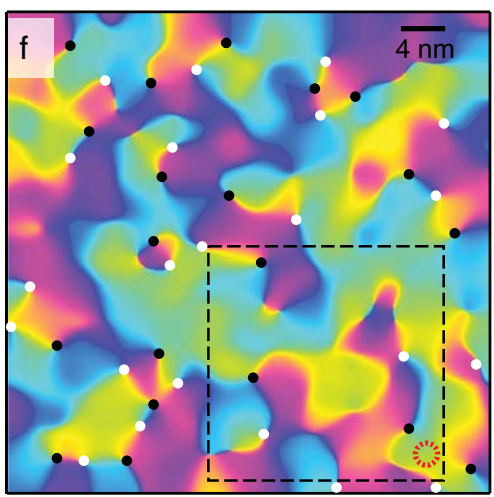

4 nm

4 nm

0 ▣ $2\pi$

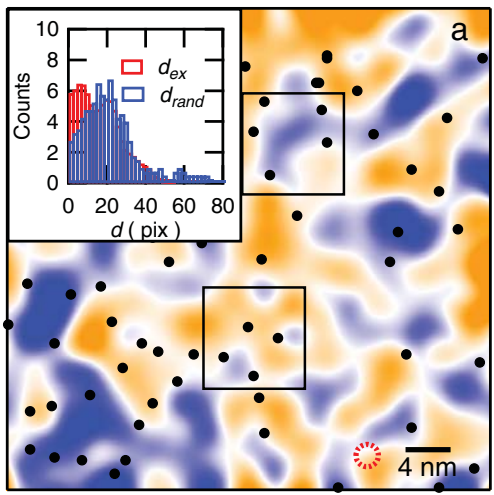

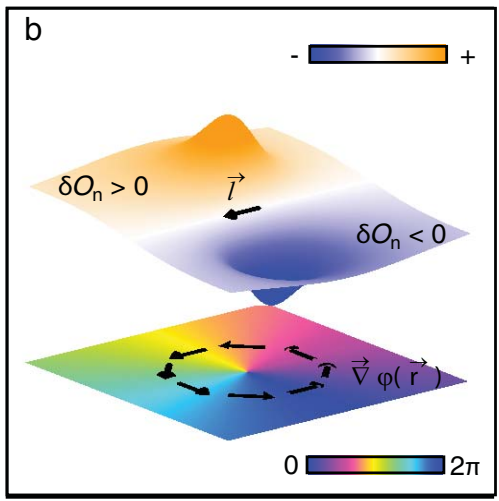

−0.0043      +0.0043

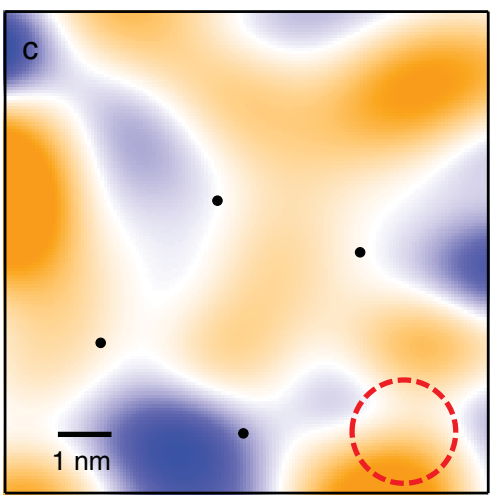

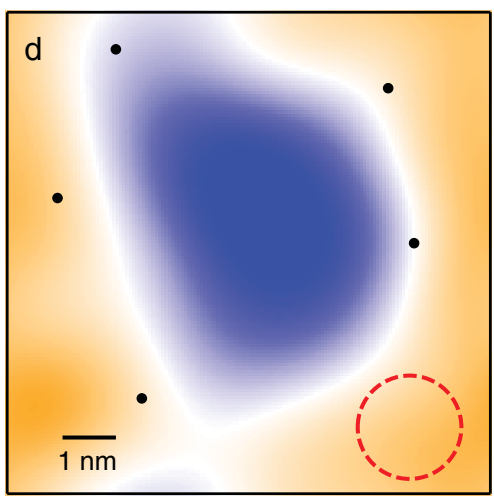

low    high        low    high

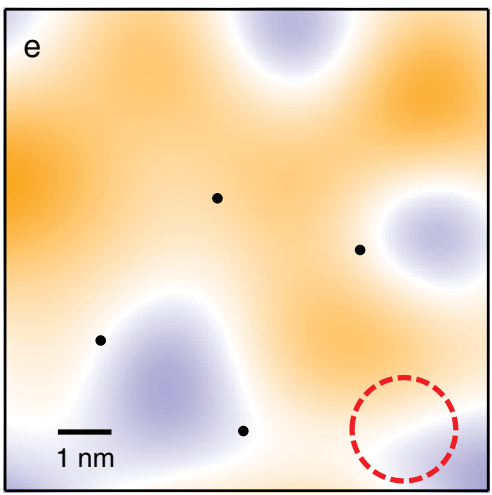

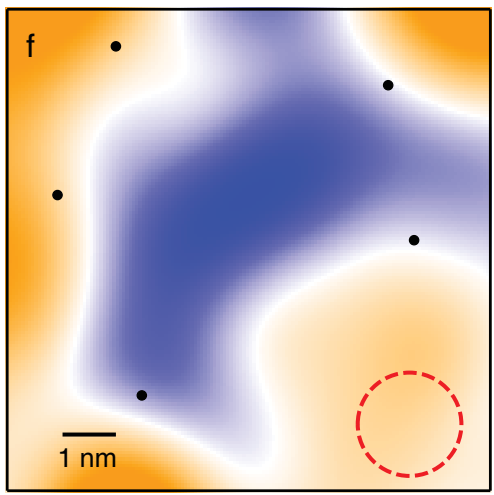

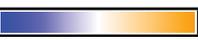

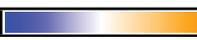

−0.0043    +0.0043      −0.0043    +0.0043

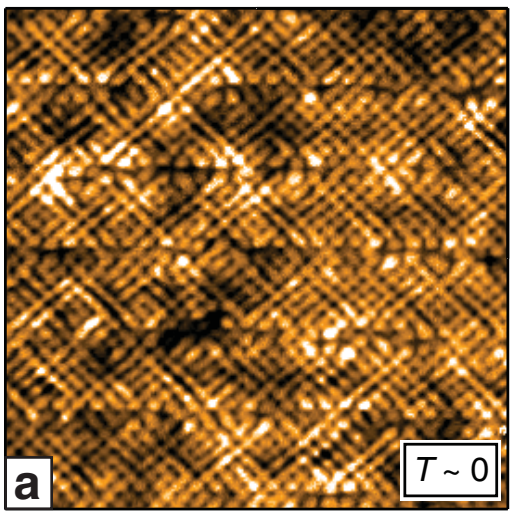

a  $T \sim 0$

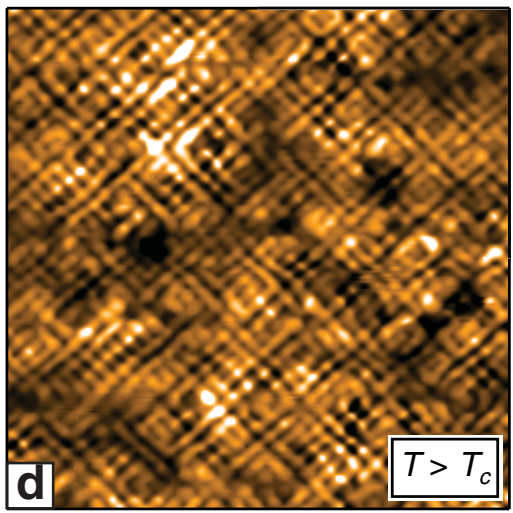

d  $T > T_c$

**I**

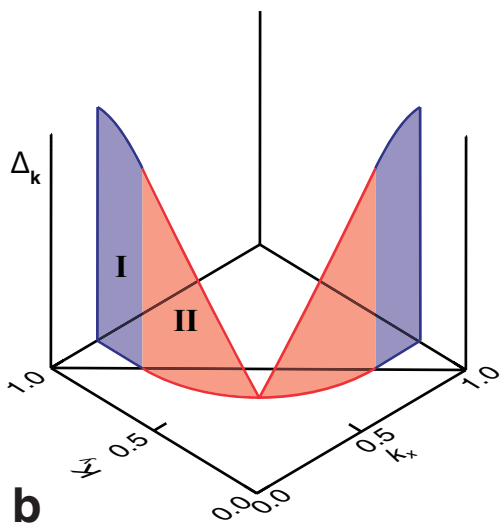

$\Delta_\mathbf{k}$

I

II

k_y  0.5

0.0 0.0

0.5  $k_x$

1.0  1.0

b

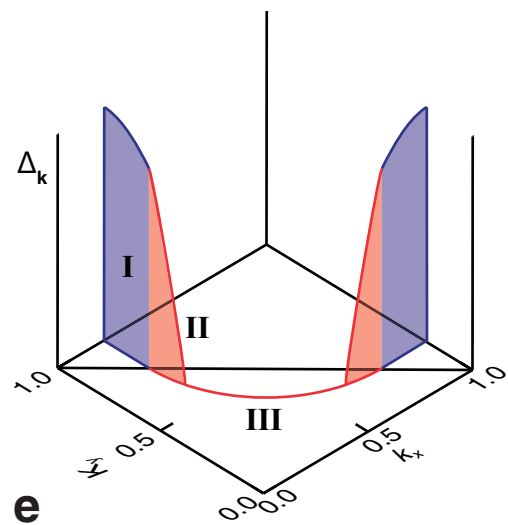

$\Delta_\mathbf{k}$

I

II

III

k_y  0.5

0.0 0.0

0.5  $k_x$

1.0  1.0

e

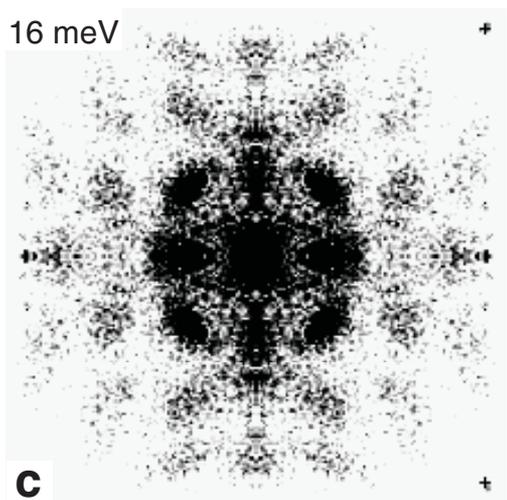

16 meV

Phase Coherent *d*-SC

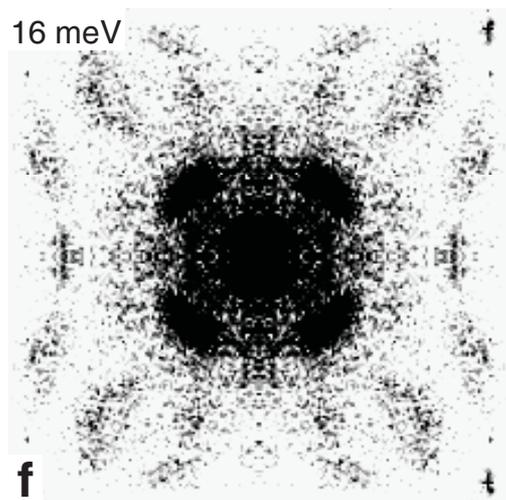

16 meV

**II**

Phase Incoherent *d*-SC